\documentclass[twocolumn,english,aps,pra,eqsecnum]{revtex4-1}
\usepackage[T1]{fontenc}
\usepackage[latin1]{inputenc}
\usepackage{bm}
\usepackage{amsmath}
\usepackage{amssymb}
\usepackage{graphicx}
\usepackage{esint}

\makeatletter

\usepackage{color}
\usepackage{bm}
\usepackage{babel}

\makeatother

\usepackage{babel}
\begin{document}

\title{Critical fluctuations in an optical parametric oscillator: when light
behaves like magnetism}

\author{Kaled Dechoum$^{1}$, Laura Rosales-Zárate$^{2}$ and Peter D. Drummond$^{2}$}

\affiliation{(1) Instituto de Física da Universidade Federal Fluminense, Boa Viagem,
24210-340, Niterói, Rio de Janeiro, Brazil \\
 (2) Centre for Quantum and Optical Science, Swinburne University
of Technology, Melbourne, Australia.}
\begin{abstract}
We study the non-degenerate optical parametric oscillator in a planar
interferometer near threshold, where critical phenomena are expected.
These phenomena are associated with non-equilibrium quantum dynamics
that are known to lead to quadrature entanglement and squeezing in
the oscillator field modes. We obtain a universal form for the equation
describing this system, which allows a comparison with other phase
transitions. We find that the unsqueezed quadratures of this system
correspond to a two-dimensional XY type model with a tricritical Lifshitz
point. This leaves open the possibility of a controlled experimental
investigation into this unusual class of statistical models. We evaluate
the correlations of the unsqueezed quadrature using both an exact
numerical simulation and a Gaussian approximation, and obtain an accurate
numerical calculation of the non-Gaussian correlations. 
\end{abstract}
\maketitle

\section{Introduction}

Non-equilibrium pattern-formation occurs in many physical systems,
giving rise to the emergence of order on macroscopic scales~\cite{bowman}.
The theory of hydrodynamics is a paradigm for understanding these
phenomena, and is applicable to many branches of physics, chemistry,
biology, astrophysics, and other sciences~\cite{Pattern formation}.
This is often applied to many body systems subject to nonlinear coupling
in a dissipative environment with external fluxes. In physics, one
of the most studied hydrodynamic effects is the theory of fluid flows
near the Rayleigh-B\'enard instability instability~\cite{bowman}.
The simplest nontrivial model for fluid convection that displays pattern
formation due to convective instabilities was derived by Swift and
Hohenberg~\cite{SHeq}, where nonlinear coupling of fluctuations
was included to demonstrate the failure of mean field theory for critical
exponents.

Here, we treat pattern formation and universality in a paradigmatic
non-equilibrium \emph{quantum} system: the non-degenerate parametric
oscillator (OPO). Theories of a similar nature have been applied to
non-equilibrium spatially extended structures in lasers and other
related systems, with an emphasis on universal behavior of phase-transitions,
pattern formation and self-organization~\cite{Graham-Haken}. Yet
down-conversion in a non-degenerate parametric system can display
new possibilities not found in these simpler cases. In particular,
one can have entanglement and EPR paradoxes~\cite{ReidEPR,Ou}.

In the present paper, we investigate a new type of critical point
phase transition in this non-equilibrium quantum system in order to
understand the universality class. In a type II OPO, there are two
down-converted fields with orthogonal polarization. Hence, the order
parameter is a complex or vector field in two dimensions. The two
components of this vector field are associated with the polarization
degrees of freedom of the down-converted radiation field. This is
a quantum system driven to a phase transition far from thermal equilibrium.
It is also known to display strong quantum entanglement and EPR correlations~\cite{ReidEPR}
in the case where there are two correlated output modes. We wish to
understand the behavior of this phase-transition using a first-principles
analysis~\cite{ReidCorrelations,PM_PRA41} of the relevant master
equation.

Experimentally, the OPO is now a mature technology with both commercial
and fundamental applications. Following initial theoretical predictions~\cite{YurkeIO,InputOutputTheo},
the quantum limited type I OPO was investigated experimentally by
Wu et al~\cite{Wu}, demonstrating quantum squeezing. Later the type
II case, in a triply resonant cavity, was used to experimentally demonstrate
continuous variable EPR correlations~\cite{Ou}, also originally
predicted theoretically~\cite{ReidCorrelations,PM_PRA41}. These
initial experimental investigations were in few mode devices. Both
below and above threshold experiments have been carried out, as close
as $\pm1\%$ of the critical point~\cite{Feng2003,Villar2007,LauratHal,LauratFabre,OPOII,Keller},
confirming predictions of thresholds and conversion efficiency. Operation
at the critical point results in low-frequency critical fluctuations
and non-Gaussian behavior \cite{D'Auria}.

Spatially extended pattern formation in a degenerate or type I OPO
has also been analyzed previously~\cite{DegOPO}. It is related to
the Lifshitz phase transition~\cite{Lifshitz}. This is a model used
to describe the phase transition to a modulated magnetic phase~\cite{Michelson,Hornreich79}.
In this simplest case one has a two dimensional, planar system with
a scalar order parameter~\cite{Planar}, which has known universality
properties. A Swift Hohenberg equation was derived for spatially extended
nondegenerate type I OPOs with flat end mirrors \cite{MorcilloSH},
but ignoring fluctuations. OPO experiments have been recently extended
to these types of multi-mode devices~\cite{Fabre,Ducci}, with type
II as well as type I parametric downconversion. The experimental situation
is that while multimode experiments have mostly used confocal mirrors,
the first type II multimode planar mirror experiments have been carried
out~\cite{Ducci}.

Studies of the Swift Hohenberg equation in the neighborhood of the
critical point, the Lifshitz point, as well as pattern formation for
lasers have been discussed ~\cite{LegaLifshitzPNonOpt}. The Lifshitz
point is similar to the tricritical point \cite{Hornreich}. Tricritical
points occur in different physical systems. They correspond to a point
in a phase-diagram where two lines of ordinary critical points meet
and terminate \cite{TP}. For critical points and tricritical points
there are two critical dimensions. The \emph{upper} critical dimension
refers to the one above which the critical exponents have classical
values \cite{Hornreich}. The \emph{lower} critical dimension refers
to the smallest dimension for which there is a true phase-transition,
due to increasing fluctuations as the dimension is decreased.

Systems like these can entangle large numbers of modes, with quantum
entanglement increasing as threshold is approached. Currently, experiments
near the critical point are sensitive to classical fluctuations~\cite{D'Auria}
and heating effects~\cite{Ducci}. With improved stabilization methods,
we expect that these technical problems can be overcome. However,
noise due to quantum fluctuation effects will remain.

Theoretically, the usual approach for two-dimensional non-equilibrium
problems is the Landau-Ginzburg (LG) equation~\cite{Staliunas},
which was first used for understanding superconductivity near threshold.
Normally, the LG equation is derived using approximations like adiabatic
elimination, and is taken as an effective equation for the physical
system. There is now an increased interest in extended spatial and
multi-mode structures~\cite{Lugiato,Fabre} in the quantum optical
parametric oscillator (OPO)~\cite{Peter,Wu}, due to availability
of multiple transverse mode cavities~\cite{Marte} and quantum imaging
control~\cite{Pointer}. It is important in these cases to understand
how quantum noise enters the dynamical equations~\cite{Swain}.

We show that planar type-II down-conversion creates a non-equilibrium
system with a similar general type of symmetry to the Berezinskii~\cite{Berezinskii}
and Kosterlitz-Thouless (BKT) model~\cite{K-T}, but with an isotropic
Lifshitz point, first studied in magnetic systems by Hornreich et
al~\cite{Hornreich}. Physically, this is not a classical fluid or
magnetic system, but rather is a quantum system, driven into a non-equilibrium
critical point~\cite{CDD} far from thermal equilibrium. The non-degenerate
planar OPO is fundamentally different to the usual BKT model, having
a quartic rather than quadratic momentum-dependence in the linear
response function. This places it in a similar category to the Swift-Hohenberg
model and next-nearest neighbor lattice models. Our model can also
display strong EPR entanglement and other nonclassical properties
in addition to the Lifshitz point behavior. The parametric model therefore
provides a novel path to the investigation of unusual classical and
quantum noise effects.

Phase transitions with this general type of symmetry are continuous
yet break no symmetries. There is also no ferromagnetism in this system,
as this is prohibited by the Mermin-Wagner theorem~\cite{Mermin-Wagner}.
Berezinskii predicted a new type of phase with correlations that decay
slowly with distance with a power law. The phase transition for a
ferromagnetic phase is prevented by the appearance of vortex and anti-vortex
pairs. Many quantum phase transitions in two dimensions belong to
this class. The continuous version of the XY or Ising models are often
used to model systems that possess order parameters with a symmetry
of this type, e.g. superfluid helium, liquid crystals, two dimensional
Bose Einstein condensate (BEC), and others. While having the same
order-parameter and space dimension as this case, we show that the
type II OPO has a different universality class.

Our approach is to start from the usual master equation model that
describes a type II OPO with transverse modes. Here we consider that
the OPO is non-degenerate in polarization. We then map the coupled
Heisenberg equations into the positive P-representation~\cite{PosP}
of the density matrix. This choice is made because the positive P-representation
allows us to \emph{exactly} map the density matrix evolution into
an equivalent stochastic equation.

We organize this paper in the following way. First we describe the
Hamiltonian model and derive the equation of motion for this open
system. Next, we use the positive-P representation for mapping these
equation into a Langevin type, which can either be treated numerically
or via analytic approximations. We will mostly describe the unsqueezed
quadratures of the field, which have a large similarity with the corresponding
magnetic system. At this point we can recognize the universality class.
We give both a Gaussian approximation to the correlation functions
of the unsqueezed quadratures, and high-precision numerical simulations
of the non-Gaussian corrections. We show that, even though the system
is far below its upper critical dimension, the non-Gaussian character
of the fluctuations is relatively small, and the intensity fluctuations
are nearly factorizable.

In a following paper we will focus our attention on the squeezed field
quadratures, to give a spatial map of quantum squeezing and Einstein-Podolsky-Rosen
(EPR) entanglement.

\section{The model}

The system of interest comprises an optically driven planar Fabry-Perot
cavity or interferometer with a nonlinear medium that possess a parametric
nonlinearity. The nonlinear crystal is cut to give a type II phase
matching, that couples a pump field to two down-converted fields having
an orthogonal polarization. The cavity is pumped with a spatially
extended coherent light with frequency $\omega_{0}$, with a transverse
spatial profile. In the simplest case we consider this pump to be
a plane wave.

\subsection{Quantum Hamiltonian}

The outgoing down-converted light, amplified inside the cavity, develops
structures and patterns due to diffraction, nonlinear coupling, and
detuning between the wavelength of the down-converted field and the
cavity size. This depends on the modal decomposition of this cavity.
The mirrors have parameters that can be controlled by experimentalists.
The tunable parameters include the reflection coefficient for each
mode, and the cavity detunings for each mode.

Our model for this system is similar to many earlier treatments of
driven nonlinear optical cavities~\cite{Haken,InputOutputTheo,ReidCorrelations,Lugiato,Ritsch,GardinerQNoise,InputOutputBooks,PeterBook}.
It includes a linear coupling between the external electromagnetic
field modes and the internal cavity modes, owing to a partially transmitting
mirror. The cavity and mirror parameters determine both the coupling
to the driving field and the decay rate of the cavity or interferometer.
We note that for a low-Q device, it is important to use non-orthogonal
quasi-modes~\cite{Ritsch}. Here, we assume the opposite case of
a thin, high-Q extended planar cavity, so that the external modes
are simply plane-wave modes.

The quantum Hamiltonian in the interaction picture has four main terms
that can be summarized by the following expression: 
\begin{equation}
\hat{H}=\hat{H}_{free}+\hat{H}_{int}+\hat{H}_{pump}+\hat{H}_{res}\,,\label{eq:QHam}
\end{equation}
where there are three fundamental Bose fields, $\hat{A}_{i}(\bm{x},t)$
for $i=0,2$. The boson fields are the transverse internal modes of
the planar cavity, which is assumed to have a single longitudinal
mode in the direction normal to the mirrors. We note that this planar
polariton field model is also used in the theory of a polaritonic
BEC~\cite{CarusottoRMP}. The boson fields have two orthogonal polarizations,
$i=1,2$, and obey the usual equal time commutation relation, 
\[
\left[\hat{A}_{i}(\bm{x},t),\hat{A}_{j}^{\dagger}(\bm{x}',t)\right]=\delta_{ij}\delta^{2}\left(\bm{x}-\bm{x}'\right).
\]
Here $\delta^{2}\left(\bm{x}-\bm{x}'\right)$ is a Dirac delta function
in the two-dimensional transverse plane of $\bm{x}=(x,y)$. The cavity
fields are defined in terms of polaritonic annihilation and creation
operators as: $\hat{A}_{i}(\bm{x},t)=\sum_{\bm{k}}e^{i\bm{k}\cdot\bm{x}}\hat{a}_{i}(\bm{k},t)/L,$
where $\hat{a}_{i}(\bm{k},t)$ represents the annihilation operator
for a free polariton mode with transverse momentum $\bm{k}$, and
the summation is over a set of $M$ discrete modes with periodic boundary
conditions, on a large area $L^{2}$.

The free evolution Hamiltonian that accounts for diffraction inside
the planar cavity is: 
\begin{equation}
\hat{H}_{free}=\sum_{i=0}^{2}\hbar\int d^{2}\bm{x}\,\hat{A}_{i}^{\dagger}\left[\omega_{i}-\frac{v_{i}^{2}}{2\omega_{i}}\nabla^{2}\right]\hat{A}_{i}.
\end{equation}
This Hamiltonian describes a planar cavity with intra-cavity resonant
frequencies $\omega_{i}$ and group velocities $v_{i}$ for the three
field envelopes. The resonant frequencies have the relation $\omega_{0}\approx\omega_{1}+\omega_{2}$,
where $\omega_{0}$ is the fundamental mode, and $\omega_{1}$ and
$\omega_{2}$ stand for the down-converted light. The case of spherical
mirrors can also be analyzed in a similar way~\cite{Marte}, but
is not treated here. We suppress the field space-time arguments to
obtain more compact expressions in the integrals. The two-dimensional
Laplacian is, as usual, $\nabla^{2}=\partial^{2}/\partial x^{2}+\partial^{2}/\partial y^{2}$.
A one-dimensional system can be treated by simply dropping one of
the dimensions.

The interaction Hamiltonian representing the coupled modes inside
a crystal with $\chi^{(2)}$ nonlinearity is given by~\cite{Hillery}:
\begin{equation}
\hat{H}_{int}=i\hbar\int d^{2}\bm{x}\left[\chi\hat{A}_{0}\hat{A}_{1}^{\dagger}\hat{A}_{2}^{\dagger}-\chi^{*}\hat{A}_{0}^{\dagger}\hat{A}_{1}\hat{A}_{2}\right].
\end{equation}
This term may represent a photon of frequency $\omega_{0}$ converted
in two photons of distinct frequencies $\omega_{1}$ and $\omega_{2}$,
or photons with orthogonal polarizations, or both.

From now on, for definiteness, labels 1 and 2 stand for polarizations,
and we shall focus on the degenerate frequency case with non-degenerate
polarization. For dimensional reasons, $\chi\propto\chi^{(2)}/\sqrt{\ell}$,
where $\chi^{(2)}$ is the Bloembergen nonlinear polarizability coefficient
and $\ell$ is the intracavity longitudinal mirror spacing.

Following standard input-output theory derivations~\cite{InputOutputTheo,YurkeIO,GardinerQNoise,InputOutputBooks,PeterBook},
the Hamiltonian term associated with the input laser pumping in a
rotating frame at frequency $\omega_{L}$ is: 
\begin{equation}
\hat{H}_{pump}=i\hbar\int d^{2}\bm{x}\left[{\cal E}^{*}(\bm{x})e^{2i\omega_{L}t}\hat{A}_{0}-{\cal E}(\bm{x})e^{-2i\omega_{L}t}\hat{A}_{0}^{\dagger}\right].
\end{equation}
While it is possible to choose any shape carrying spatial structure
for the input pump, here for simplicity we will assume a plane wave
input. The reservoir Hamiltonian is assumed to have the structure:
\begin{equation}
\hat{H}_{res}=\sum_{i=0}^{2}\int d^{2}\bm{x}\,\left[\hat{\Gamma}_{j}^{\dagger}\hat{A}_{j}+\hat{\Gamma}_{j}\hat{A}_{j}^{\dagger}\right]+\hat{H}_{res}^{0}\,.
\end{equation}
Hence, there are local coupling terms to independent external free-field
reservoirs for each polarization and each spatial mode. The external
fields are described by a free Hamiltonian $\hat{H}_{res}^{0}$.

\subsection{Master equation}

The non-unitary evolution of the system comes from the coupling between
the cavity modes and the output modes. This can be treated as a quantum
Markovian process that simulates a bath interaction. We carry out
this calculation in a type of interaction picture so that the interaction
picture operators evolve according to a reference Hamiltonian $\hat{H}_{0}$,
given by: 
\begin{equation}
\hat{H}_{0}=\hbar\sum_{j}\int d^{2}\bm{x}\,\omega_{j}^{0}\hat{A}_{j}^{\dagger}\hat{A}_{j}\,,
\end{equation}
where the reference frequencies $\omega_{1}^{0}$ are chosen so that:
\begin{align}
\omega_{1}^{0} & =\omega_{L}+\epsilon\approx\omega_{1},\nonumber \\
\omega_{2}^{0} & =\omega_{L}-\epsilon\approx\omega_{2},\nonumber \\
\omega_{0}^{0} & =2\omega_{L}\approx\omega_{0}.
\end{align}
We note that while the choice of $\omega_{L}$ is determined by the
pump frequency, the choice of $\epsilon$ is arbitrary, as long as
the Markovian approximation is still valid. Our choice of reference
Hamiltonian is not determined by the intracavity frequencies $\omega_{i}$,
which leads to detuning terms occurring in the resulting equations
of motion. This allows us to have some freedom of choice in defining
the detuning parameters. The choice will be made definite in the following
sections. Following standard techniques for deriving the master equations~\cite{GardinerQNoise,Haken,Carmichael},
we can write the master equation for the density operator in the generalized
Lindblad form: 
\begin{equation}
\frac{\partial\hat{\rho}}{\partial t}=\frac{1}{i\hbar}\left[\hat{H},\hat{\rho}\right]+\sum_{i=0}^{2}\gamma_{i}{\cal L}_{i}\left[\hat{\rho}\right],
\end{equation}
where the dissipative Liouville super-operator, 
\begin{equation}
{\cal L}_{i}\left[\hat{\rho}\right]=\int d^{2}\bm{x}\left[2\hat{A}_{i}\hat{\rho}\hat{A}_{i}^{\dagger}-\hat{\rho}\hat{A}_{i}^{\dagger}\hat{A}_{i}-\hat{A}_{i}^{\dagger}\hat{A}_{i}\hat{\rho}\right],
\end{equation}
describes the output coupling of the $i$-th intracavity mode with
the external bath.

We note that, although we focus on the master equation approach here,
it is sometimes useful to write the time evolution in a complementary
quantum Langevin formulation. This formulation is given in Appendix
A.

\section{Stochastic equations in the positive-P representation\label{sec:IIIPositiveP}}

As nonlinear operator equations are not generally soluble, it is more
manageable to map the operator equations into c-number form. In this
approach, a master equation is transformed into a positive-definite
Fokker-Planck equation using operator identities that map the operator
terms in the master equation into differential operators~\cite{Carmichael,GardinerQNoise,PeterBook}.
To do this we have to use a phase-space representation of the master
equation.

Phase-space representations in a classical phase-space do not give
a positive-definite equation. An example is the Wigner representation~\cite{Wigner}.
While this is exact, it is not able to be mapped into Langevin equations,
unless one either truncates or uses higher order noise~\cite{Higher-order}.
One can also linearize the Hamiltonian and obtain an approximate Wigner
diffusion~\cite{CDD}. Another approach is the Husimi Q-function~\cite{Husimi},
where, in order to obtain a positive Fokker-Planck equation, an unphysical
constraint on the phase-space trajectories has to be used~\cite{ZambriniQ}.

Here we wish to have the ability to treat non-equilibrium structures
without restrictions. For this purpose, the most useful representation
is the positive-P representation~\cite{PosP}, which is an extension
of the Glauber-Sudarshan P-representation~\cite{Glauber} into a
phase-space of double the classical dimensions. Unlike the P-representation,
which is singular for nonclassical states, the positive-P representation
is well-defined, positive and non-singular for any quantum state.
This approach allows us to map the density matrix equation into a
Fokker-Planck equation on a non-classical phase-space. In the positive-P
representation stochastic averages give normally-ordered quantum expectation
values. A brief description is given in Appendix B.

The stochastic field partial differential equations are given by an
extension of our earlier work~\cite{Planar}: 
\begin{eqnarray}
\frac{\partial A_{0}}{\partial t} & = & -\tilde{\gamma}_{0}A_{0}+{\mathcal{E}}(\bm{x})-\chi^{*}A_{1}A_{2}+\frac{iv_{0}^{2}}{2\omega_{0}}\nabla^{2}A_{0}\,,\nonumber \\
\frac{\partial A_{1}}{\partial t} & = & -\tilde{\gamma}_{1}A_{1}+\chi A_{0}A_{2}^{+}+\frac{iv_{1}^{2}}{2\omega_{1}}\nabla^{2}A_{1}+\sqrt{\chi A_{0}}\xi_{1}\,,\nonumber \\
\frac{\partial A_{2}}{\partial t} & = & -\tilde{\gamma}_{2}A_{2}+\chi A_{0}A_{1}^{+}+\frac{iv_{2}^{2}}{2\omega_{2}}\nabla^{2}A_{2}+\sqrt{\chi A_{0}}\xi_{2}\,.\quad\label{eq:+P-equations}
\end{eqnarray}
The three equations that correspond to the hermitian conjugate fields,
$A_{i}^{+}$, are obtained by conjugating the constant terms, and
replacing stochastic and noise fields so that: $A_{i}\rightarrow A_{i}^{+}$
and $\xi_{i}\rightarrow\xi_{i}^{+}$, where $\xi_{i}$ and $\xi_{i}^{+}$
are independent Gaussian complex noises. These are equivalent in the
mean to the conjugated Heisenberg equations, but are independent c-number
equations. They are not conjugate in every realization. We also note
that $A_{i}$ and $A_{i}^{+}$ are six independent, complex c-number
fields.

The stochastic fields $\xi_{k}$ that describe the quantum noise are
complex and Gaussian, whose non-vanishing correlations are: 
\begin{eqnarray}
\left\langle \xi_{1}(\bm{x},t)\xi_{2}(\bm{x}',t')\right\rangle  & = & \delta^{2}(\bm{x}-\bm{x}')\delta(t-t')\nonumber \\
\left\langle \xi_{1}^{+}(\bm{x},t)\xi_{2}^{+}(\bm{x}',t')\right\rangle  & = & \delta^{2}(\bm{x}-\bm{x}')\delta(t-t')\,\,.
\end{eqnarray}
This means that \ensuremath{\xi_{k}(\bm{x},t)}, \ensuremath{\xi_{k}^{+}(\bm{x},t)}
represent four independent, delta-correlated, complex c-number Gaussian
stochastic fields with zero mean. They are completely characterized
by the specified correlations. From the stochastic equations (\ref{eq:+P-equations}),
it is clear that the amplitude of the stochastic fluctuations that
act on the converted modes depend on the pump field dynamics. A brief
discussion of the noises is given in Appendix B.

\subsection{Critical driving field}

As a first investigation, we treat the classical approximation, which
has also been analyzed in some earlier work~\cite{MorcilloSH,Lugiato,Santagiustina,Zambrini,Izus,Ref3}.
Here one assumes that all noise is negligible, so that $A_{i}^{+}=A_{i}^{*}$,
which gives equations in the form: 
\begin{align}
\frac{\partial A_{0}}{\partial t} & =-\tilde{\gamma}_{0}A_{0}+{\mathcal{E}}(\bm{x})-\chi^{*}A_{1}A_{2}+\frac{iv_{0}^{2}}{2\omega_{0}}\nabla^{2}A_{0}\,,\nonumber \\
\frac{\partial A_{i}}{\partial t} & =-\tilde{\gamma}_{i}A_{i}+\chi A_{0}A_{3-i}^{*}+\frac{iv_{i}^{2}}{2\omega_{i}}\nabla^{2}A_{i}\,.
\end{align}
The phases of ${\mathcal{E}}$ and $\chi$ are essentially arbitrary,
as they depend on the phase definition for the field amplitudes $A_{i}$,
which in turn depend on arbitrary mode phases. We will use this freedom
later on to simplify the equations. If in addition, we assume that
the input is a plane-wave, and we ignore possible spatial instabilities,
diffraction can be neglected as well. A steady-state result involves
setting the time-derivatives to zero, so: 
\begin{align}
A_{0} & =\left(\mathcal{E}-\chi^{*}A_{1}A_{2}\right)/\tilde{\gamma}_{0}\,,\nonumber \\
A_{i} & =\chi A_{0}A_{3-i}^{*}/\tilde{\gamma}_{i}\,\,.\label{eq:Steady-state}
\end{align}
Hence, the defining equation for a steady-state is: 
\begin{equation}
A_{1}A_{2}^{*}=\frac{A_{1}A_{2}^{*}\left|\chi A_{0}\right|^{2}}{\tilde{\gamma}_{1}\tilde{\gamma}_{2}^{*}}\,.
\end{equation}
We can always choose the interaction picture detuning $\epsilon$
so that $\tilde{\gamma}_{1}\tilde{\gamma}_{2}^{*}=\bar{\gamma}^{2}$
is real, otherwise the above threshold solutions will be oscillatory
rather than stable. There are two types of steady-state solution.
Either $A_{1}A_{2}^{*}=0$, or else $\left|\chi A_{0}\right|=\bar{\gamma}\,$.
The first is called a below-threshold solution, the second an above-threshold
solution. There is a driving field where both the solutions coincide,
at a critical pump intensity of: 
\begin{equation}
\left|\mathcal{E}_{c}\right|^{2}=\bar{\gamma}^{2}\left|\tilde{\gamma}_{0}/\chi\right|^{2}\,.\label{eq:critical}
\end{equation}
More generally, suppose we are in the above-threshold regime. Using
Eq. (\ref{eq:Steady-state}), and defining $A_{1}^{*}A_{1}=I_{1}$,
one obtains: 
\begin{equation}
A_{0}=\left(\mathcal{E}-\left|\chi^{2}\right|I_{1}A_{0}/\tilde{\gamma}_{2}\right)/\tilde{\gamma}_{0}\,.
\end{equation}
On re-arranging the equation, and using the above-threshold solution
$\left|\chi A_{0}\right|=\bar{\gamma}\,$, this result becomes: 
\begin{equation}
\left|\chi A_{0}\right|^{2}=\bar{\gamma}^{2}=\frac{\left|\tilde{\gamma}_{2}\chi\mathcal{E}\right|^{2}}{\left|\tilde{\gamma}_{0}\tilde{\gamma}_{2}+\left|\chi^{2}\right|I_{1}\right|^{2}}\,.
\end{equation}
The roots of the resulting quadratic are: 
\begin{equation}
I_{1}=\frac{1}{\left|\chi^{2}\right|}\left[-z'\pm\sqrt{\left|\chi\mathcal{E}\right|^{2}+\left(z'\right)^{2}-\left|z\right|^{2}}\right]\,,
\end{equation}
where $z=z'+iz^{\prime\prime}=\tilde{\gamma}_{0}\tilde{\gamma}_{2}$.

Solutions with negative intensities are unphysical. There is a positive,
above-threshold solution if $\left|\chi\mathcal{E}\right|>\left|z\right|$,
which gives an identical critical field to Eq. (\ref{eq:critical}).
For $\left|\mathcal{E}\right|>\mathcal{E}_{c}$ there is a transfer
of energy from the pump to the signal and idler modes, which develop
a finite mean intensity. It is the vicinity and just above this critical
point that is the main regime of interest in this paper. We note that
above threshold, further instabilities exist, including limit cycles
and spatial pattern formation~\cite{Izus}.

\section{Adiabatic elimination of the pump mode}

We now return to the full quantum behavior given by the stochastic
equations obtained above. One limit that has an especially simple
behavior is found in the case of a rapidly decaying pump mode. We
can treat this by means of an adiabatic elimination procedure. Assuming
that $\tilde{\gamma}_{0}\gg\tilde{\gamma}_{1}\,\simeq\,\tilde{\gamma}_{2}$,
and that ${\cal {E}}$ is spatially uniform (that is, we are neglecting
pump diffraction), we can perform an adiabatic elimination by using
the stationary solution for the pump mode, so that 
\begin{equation}
A_{0}=\bar{A}_{0}\equiv\frac{{\cal E}-\chi^{*}A_{1}A_{2}}{\tilde{\gamma}_{0}}\,.
\end{equation}

\subsection{Signal and idler equations}

The resulting equations for the down-converted modes - often called
the signal and idler equations - are, for $i=1,2$: 
\begin{align}
\frac{\partial A_{i}}{\partial t}(\bm{x},t) & =-\tilde{\gamma}_{i}A_{i}+\frac{\chi}{\tilde{\gamma}_{0}}\left({\cal E}-\chi^{*}A_{1}A_{2}\right)A_{3-i}^{+}+\nonumber \\
 & +\frac{iv_{i}^{2}}{2\omega_{i}}\nabla^{2}A_{i}+\sqrt{\chi\bar{A}_{0}}\xi_{i}(\bm{x},t)\,.\label{eq:Eq20}
\end{align}
We see that the main effect of detuning the pump is to reduce the
effective intra-cavity pump intensity. So far, we have treated the
case of general frequencies and group velocities. An important special
case is obtained when the two down-converted frequencies are equal.
In this case, the modes are still non-degenerate, as they can have
different polarizations. From now on, we consider this special case
for simplicity, so we take $\tilde{\gamma}_{1}=\tilde{\gamma}_{2}=\tilde{\gamma}$,
$v_{1}=v_{2}=v$ and $\omega_{1}=\omega_{2}=\omega$. This implies
that $\Delta_{1}=\Delta_{2}=\Delta$. We will also assume that $\Delta_{0}=0$,
i.e. that the pump is on-resonance with the cavity, even when the
down-converted fields may be off-resonant from their cavity resonance
frequencies.

Although these assumptions simplify the algebra, they are not essential
for our main conclusions, which mostly rely simply on the fact that
we now have a two-dimensional order parameter rather than a one-dimensional
order parameter as found in the degenerate case. We should note that
a non-zero pump detuning can excite another modes different from $\omega_{0}$.
This could give rise to nonlinear phenomena like bistabilities, nonlinear
resonances and subcritical bifurcation, especially for the case of
large detuning \cite{MorcilloSH}. In the case where the diffraction
terms are different, there will be a new term which is proportional
to the difference of the two diffraction terms, as has been studied
elsewhere \cite{MorcilloSH}. This term will be present in the equations
even for the case of zero detuning. Since we are interesting in the
universal behaviour of the system we do not include these cases, but
we point out that they can give rise to nonlinear behaviour.

\subsection{Dimensionless form}

Equation (\ref{eq:Eq20}) will now be transformed into a dimensionless
form which allows comparisons with other types of phase transitions.
First, we define the dimensionless variables $\tau=t/t_{0}$, $\bm{r}=\bm{x}/x_{0}$,
with a scaled Laplacian $\nabla_{\bm{r}}^{2}=\partial^{2}/\partial r_{1}^{2}+\partial^{2}/\partial r_{2}^{2}$.
It is useful to also define a dimensionless field $\alpha_{i}=x_{0}A_{i}$
as well. This has an intuitive interpretation as the coherent amplitude
in real space, defined relative to a physical area of $x_{0}^{2}.$
After this transformation, one obtains: 
\begin{equation}
\frac{1}{t_{0}}\frac{\partial\alpha_{i}}{\partial\tau}=-\tilde{\gamma}\alpha_{i}+\chi\bar{A}_{0}\alpha_{3-i}^{+}+\frac{iv^{2}}{2\omega x_{0}^{2}}\nabla_{\mathbf{r}}^{2}\alpha_{i}+x_{0}\sqrt{\chi\bar{A}_{0}}\xi_{i}\,.
\end{equation}
We define $\tilde{\gamma}=\gamma\left(1+i\Delta\right)$ and introduce
the dimensionless pump amplitude 
\begin{equation}
\tilde{\mu}=\frac{\chi{\cal E}}{\gamma_{0}\gamma}=\mu e^{i\phi}\,,
\end{equation}
where $\mu$ is real and positive. Since the phase of the driving
field ${\cal E}$ is arbitrary, and the equations are invariant under
phase-changes of ${\cal {E}}$, we will choose $\phi=0$ with no loss
of generality. We also define the time-scale $t_{0}=1/g\gamma$ as
the scaling time for critical slowing down, where 
\begin{equation}
g^{2}=\frac{\left|\chi\right|^{2}}{4\gamma_{0}\gamma x_{0}^{2}}\,.\label{eq:g-definition}
\end{equation}
The length scale $x_{0}$ is now chosen so that: 
\begin{equation}
x_{0}^{2}=\frac{v^{2}}{2\gamma\sqrt{g}\omega}\,.\label{eq:x-definition}
\end{equation}
Combining Eqs. (\ref{eq:g-definition}) and (\ref{eq:x-definition}),
one can write the dimensionless coupling $g$ in the form: 
\begin{equation}
g=\left(\frac{\left|\chi\right|^{2}\omega}{2\gamma_{0}v^{2}}\right)^{2/3}\,.
\end{equation}
Similarly, we introduce dimensionless complex noises $\zeta_{i}=\xi_{i}x_{0}\sqrt{t_{0}}$,
so that 
\begin{eqnarray}
\left\langle \zeta_{1}(\bm{r},\tau)\zeta_{2}(\bm{r}~',\tau')\right\rangle  & = & \delta^{2}(\bm{r}-\bm{r}~')\delta(\tau-\tau'),\nonumber \\
\left\langle \zeta_{1}^{+}(\bm{r},\tau)\zeta_{2}^{+}(\bm{r}~',\tau')\right\rangle  & = & \delta^{2}(\bm{r}-\bm{r}~')\delta(\tau-\tau').
\end{eqnarray}
Finally, defining the driving field saturation factor as: 
\begin{align}
\mu\left(\vec{\alpha}\right) & =\mu-4g^{2}\alpha_{1}\alpha_{2},\nonumber \\
\mu^{+}\left(\vec{\alpha}\right) & =\mu-4g^{2}\alpha_{1}^{+}\alpha_{2}^{+},
\end{align}
we obtain the following dimensionless form of the scaled equations
for $i=1,2$: 
\begin{align}
g\frac{\partial\alpha_{i}}{\partial\tau} & =\mu\left(\vec{\alpha}\right)\alpha_{3-i}^{+}-\left(1+i\Delta\right)\alpha_{i}\nonumber \\
 & +\sqrt{g}\left[i\nabla_{\bm{r}}^{2}\alpha_{i}+\sqrt{\mu\left(\vec{\alpha}\right)}\zeta_{i}(\bm{r},\tau)\right],\label{eq:ScaledEqs}
\end{align}
together with a conjugate equation for $\alpha_{i}^{+}$. Although
this is true in general, we are mostly interested here in the regime
of $g\ll1$, which allows us to make an expansion in powers of the
coupling. For small $g$, the classical approximation gives the leading
order term, while the diffraction and noise terms provide the next
order in an expansion in $\sqrt{g}$.

\section{Stability properties and quadrature equations}

We now wish to transform these equations into quadrature equations
that are simpler to investigate. There are very different stability
properties for the orthogonal quadratures near the critical point.
To investigate this, as a first approximation, we will ignore noise
and nonlinear terms of order $\sqrt{g}$ and smaller. The stability
of the equations near $\alpha_{i}=0$, to leading order in $g$, is:
\begin{equation}
g\frac{\partial}{\partial\tau}\left(\begin{array}{c}
\alpha_{1}\\
\alpha_{2}^{+}
\end{array}\right)=\left(\begin{array}{cc}
-\left(1+i\Delta\right) & \mu\\
\mu & -\left(1-i\Delta\right)
\end{array}\right)\left(\begin{array}{c}
\alpha_{1}\\
\alpha_{2}^{+}
\end{array}\right),
\end{equation}
which has eigenvalues $\lambda_{\pm}=-1\pm\sqrt{-\Delta^{2}+\mu^{2}}$,
and with a similar equation coupling $\alpha_{2}$ and $\alpha_{1}^{+}$.
This gives an unstable eigenvalue, leading classically to growth of
the signal and idler terms if $\mu^{2}-\Delta^{2}>1$, as expected
from the analysis in the previous section. The resulting eigenvectors,
$\vec{u}_{\pm}$, are: 
\begin{equation}
\vec{u}_{\pm}=\left(\begin{array}{c}
\mu\\
i\Delta\pm\sqrt{-\Delta^{2}+\mu^{2}}
\end{array}\right).
\end{equation}
There is clearly a line of critical points where the stability has
a continuous change at $\mu^{2}-\Delta^{2}=1$. The tricritical point
then occurs when $\mu=1,\Delta=0$, as we will show in later sections.

\subsection{Quadrature field variables}

To understand the behavior of Eq. (\ref{eq:ScaledEqs}) in the neighborhood
of the critical point, we define complex, dimensionless scaled quadrature
fields which are proportional to the critical eigenvectors, as follows~\cite{Planar,CDD}:
\begin{align}
X= & \sqrt{g}\left(\alpha_{1}+\alpha_{2}^{+}\right),\nonumber \\
X^{+}= & \sqrt{g}\left(\alpha_{2}+\alpha_{1}^{+}\right),\nonumber \\
Y= & \frac{1}{i}\left(\alpha_{1}-\alpha_{2}^{+}\right),\nonumber \\
Y^{+}= & \frac{1}{i}\left(\alpha_{2}-\alpha_{1}^{+}\right)\,.\label{eq:Quadratures_theta0}
\end{align}

Next, we consider the case $\mu\approx1\gg g^{2}\left|\alpha_{1}\alpha_{2}\right|$,
to simplify the noise term. As this is both relatively small, and
nearly constant in the neighborhood of the critical point, the resulting
terms are of higher order in $g$ than the leading terms we wish to
include. The resulting equations for these quadratures in the positive
P-representation near the critical point are: 
\begin{eqnarray}
\frac{\partial X}{\partial\tau} & = & \frac{\mu_{-}}{g}X+{\cal D}_{+}Y-\left(X^{2}+gY^{2}\right)X^{+}+\zeta_{+}.\nonumber \\
\frac{\partial X^{+}}{\partial\tau} & = & \frac{\mu_{-}}{g}X^{+}+{\cal D}_{+}Y^{+}-\left(X^{+2}+gY^{+2}\right)X+\zeta_{+}^{*}.\nonumber \\
g\frac{\partial Y}{\partial\tau} & = & -\mu_{+}Y+{\cal D}_{-}X-g\left(X^{2}+gY^{2}\right)Y^{+}-i\sqrt{g}\zeta_{-}.\nonumber \\
g\frac{\partial Y^{+}}{\partial\tau} & = & -\mu_{+}Y^{+}+{\cal D}_{-}X^{+}-g\left(X^{+2}+gY^{+2}\right)Y-i\sqrt{g}\zeta_{-}^{*}.\nonumber \\
\label{eq:EqsQuad}
\end{eqnarray}
Here we have defined a modified Laplacian and driving term as: 
\begin{align}
{\cal D}_{\pm} & =\pm\frac{\Delta}{\sqrt{g}}\mp\nabla_{\bm{r}}^{2}\nonumber \\
\mu_{\pm} & =\mu\pm1.
\end{align}
and new Gaussian noise terms according to: 
\begin{align}
\zeta_{\pm} & =\zeta_{1}\pm\zeta_{2}^{+}=\left(\zeta_{2}\pm\zeta_{1}^{+}\right)^{*},
\end{align}
where we have used the result of Eq. (\ref{eq:conjugate-noise}).
This shows that the noise terms driving the $X,\:X^{+}$ fields are
conjugate, while those driving the $Y,\:Y^{+}$ fields change sign
on conjugation, so that $Y,\:Y^{+}$ will not remain conjugate during
time-evolution.

At this stage, we can make the following remarks. The stochastic quadrature
fields $X,X^{+}$ are both complex fields, so they have four degrees
of freedom between them, and similarly for $Y,\:Y^{+}$. They have
a correspondence with non-Hermitian operator fields $\hat{X},\:\hat{X}^{\dagger}$
and $\hat{Y},\:\hat{Y}^{\dagger}$. In general, $Y,\:Y^{+}$ are not
complex conjugate except in the mean, and neither are $X,\:X^{+}$,
since they are driven by the $Y,\:Y^{+}$ fields. However, as we will
show, this picture simplifies when one considers an expansion near
the critical point.

\subsection{Critical point adiabatic elimination}

We can now perform a second type of adiabatic elimination which is
valid in the neighborhood of the critical point. This takes into account
the fact that the fluctuations in the $X$ quadrature become very
slow near threshold, while the $Y$ quadrature still responds on the
fast relative time scale $1/\gamma$. Formally, we can drop terms
of ${\cal O}(\sqrt{g})$ where $g\ll1$, and approximate equations
(\ref{eq:EqsQuad}) as follows~\cite{Planar}: 
\begin{eqnarray}
\frac{\partial X}{\partial\tau} & = & -\left(\frac{1-\mu}{g}\right)X+\left(\frac{\Delta}{\sqrt{g}}-\nabla_{\bm{r}}^{2}\right)Y-X^{2}X^{+}+\zeta_{+}\,,\nonumber \\
\frac{\partial X^{+}}{\partial\tau} & = & -\left(\frac{1-\mu}{g}\right)X^{+}+\left(\frac{\Delta}{\sqrt{g}}-\nabla_{\bm{r}}^{2}\right)Y^{+}-X^{+2}X+\zeta_{+}^{*}\,,\nonumber \\
0 & = & -\left(1+\mu\right)Y+\nabla_{\bm{r}}^{2}X\,,\nonumber \\
0 & = & -\left(1+\mu\right)Y^{+}+\nabla_{\bm{r}}^{2}X^{+}.
\end{eqnarray}
Here we have considered that the noise term of the quadrature variables
$Y$, $Y^{+}$ has been neglected since it scales as $\sqrt{g}$ and
we are considering the limit $g\ll1$. If this limit is not taken,
the noise term would appear in the above equations. To lowest order
in $g$, we can eliminate the fast or non-critical quadrature $Y,\:Y^{+}$
variables by setting: 
\begin{align}
Y^{(+)} & =\frac{\nabla^{2}X^{(+)}}{1+\mu}\,,
\end{align}
which gives the result that: 
\begin{eqnarray}
\frac{\partial X}{\partial\tau} & = & \mathcal{\tilde{D}}X-X^{2}X^{+}+\zeta_{+}\,,\nonumber \\
\frac{\partial X^{+}}{\partial\tau} & = & \mathcal{\tilde{D}}X-\left(X^{+}\right)^{2}X+\zeta_{+}^{*}.\label{eq:main}
\end{eqnarray}
Here we have introduced a linear differential operator that describes
the linear gain, loss and diffraction terms: 
\begin{equation}
\mathcal{\tilde{D}}\equiv\mathcal{\tilde{D}}_{r}=-\eta_{1}+\eta_{2}\nabla_{\bm{r}}^{2}-\eta_{3}\nabla_{\bm{r}}^{4}\label{eq:tildeD}
\end{equation}
where we have defined the following parameters: 
\begin{align}
\eta_{1} & =\left(\frac{1-\mu}{g}\right),\nonumber \\
\eta_{2} & =\frac{\Delta}{\left(1+\mu\right)\sqrt{g}},\nonumber \\
\eta_{3} & =\frac{1}{1+\mu}.\label{eq:etas}
\end{align}
We note that since the non-conjugate variables $Y,\:Y^{+}$ do not
appear in these equations, it follows that variables $X,\:X^{+}$
will remain conjugate if they are conjugate initially, as for example
in an initial thermal or coherent state. Hence, to leading order,
we can set $X^{+}=X^{*}$ near the critical point, and write one complex
equation for the critical quadratures, which is valid near threshold:
\begin{eqnarray}
\frac{\partial X}{\partial\tau} & = & \mathcal{\tilde{D}}X-X\left|X\right|^{2}+\zeta_{+}\,.\label{eq:mainMX}
\end{eqnarray}
While this equation is similar to a Landau-Ginzburg equation for a
complex order parameter~\cite{LPVol9}, it is not identical, owing
to the presence of the fourth-order Laplacian term in $\mathcal{\tilde{D}}$.
In the next section we will show that the equations are similar to
a vector Swift-Hohenberg equation~\cite{Pattern formation}.

\subsection{Vector Swift-Hohenberg equations }

We will now show that these near-threshold equations are actually
coupled or vector Swift-Hohenberg equations~\cite{Pattern formation}
that represent the leading order dynamics near threshold of the down-converted
modes with the same frequency but orthogonal polarization. To demonstrate
this, it is convenient to make the following change of variables:
\begin{equation}
X_{1}=\frac{X+X^{*}}{2},\qquad X_{2}=\frac{X-X^{*}}{2i}.
\end{equation}
After this change of variables it is possible to write the equation
(\ref{eq:mainMX}) as a single vector equation, as we now show. On
inverting this equation we get: 
\begin{equation}
X=X_{1}+iX_{2},\qquad X^{*}=X_{1}-iX_{2}.\label{eq:X1_X2-1}
\end{equation}
Next we define define two real vectors $\bm{X}$ and $\tilde{\bm{\zeta}}$
as: 
\[
\bm{X}=\left(\begin{array}{c}
X_{1}\\
X_{2}
\end{array}\right),\qquad\tilde{\bm{\zeta}}=\frac{1}{2}\left(\begin{array}{c}
\zeta_{+}+\zeta_{+}^{*}\\
i\left(\zeta_{+}-\zeta_{+}^{*}\right)
\end{array}\right).
\]
Using the above expressions we can write Eq. (\ref{eq:mainMX}) as
a single vector equation of the form: 
\begin{align}
\frac{\partial{\bm{X}}}{\partial\tau} & =\tilde{\mathcal{D}}\bm{X}-\left|\bm{X}\right|^{2}\bm{X}+\tilde{\bm{\zeta}},\label{eq:cvgl}
\end{align}
where ${\bm{X}}$ is a two real component vector whose elements are
$X_{1}$ and $X_{2}$, and $\tilde{\bm{\zeta}}$ is also a two real
component Gaussian noise vector, with correlations given by: 
\begin{equation}
\left\langle \tilde{\zeta}_{i}(\bm{r},\tau)\tilde{\zeta}_{j}(\bm{r}',\tau')\right\rangle =\delta_{ij}\delta(\bm{r}-\bm{r}')\delta(\tau-\tau').
\end{equation}
\begin{figure}[htbp]
\includegraphics[width=1\linewidth]{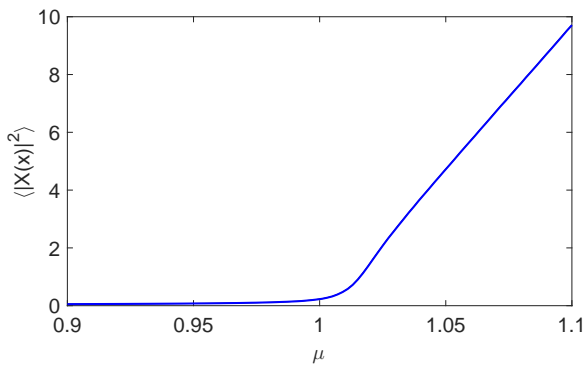} \protect\caption{Dimensionless intensity versus dimensionless pump $\mu$ in the vicinity
of the critical point. The point $\mu=1$ corresponds to $\eta_{1}=\eta_{2}=0$
and $\eta_{3}=\frac{1}{2}$. Here we have used the parameter $g=0.01$.
Results were obtained from a simulation with 300 samples on a $50\times20\times20$
numerical grid of $30000\times96\times96$ points. \label{fig:OrderParamPump} }
\end{figure}

In Fig. (\ref{fig:OrderParamPump}) we plot the modulus squared of
the two-dimensional order parameter $\bm{X}$ vs the dimensionless
pump parameter $\mu$, averaged over a transverse area of dimensionless
size $20\times20$.

Direct simulations were employed, using both a central partial difference
algorithm in the interaction picture \cite{Werner} and a fourth order
interaction picture Runge-Kutta method. Two public domain software
packages were used and checked against each other \cite{Software}.
This figure is obtained by solving the stochastic equations given
in Eq. (\ref{eq:cvgl}) at $\mu=0.9$, equilibrating for $t=50$ units
then scanning the driving field $\mu$ adiabatically (using $\mu=0.9+0.004t$)
through the critical point until $\mu=1.1$. There is a rate-independent
critical region at the threshold value of $\mu=1$, where the transition
is smooth rather than discontinuous, as it would be classically.

In the case of a one-dimensional, real order parameter, this equation
was first derived by Swift and Hohenberg~\cite{SHeq,RealSH,Pattern formation}.
Here it appears as a two-dimensional vector equation, since the order
parameter is two-dimensional. This was used to explain the convective
roll patterns generated by the Rayleigh-Bénard instability, where
the order parameter is the vertical fluid velocity. The real order
parameter case is also similar to the Ising model for magnets in two
dimensions with next nearest neighbor interactions~\cite{Yin}, where
competition between the nearest and next nearest interactions generates
a magnetic modulated phase called the Lifshitz phase~\cite{Hornreich79}.

Higher dimensional or complex order parameters, as in the present
analysis, are described using a generalized Landau-Ginzburg-Wilson
Hamiltonian~\cite{Hornreich}, where the authors also introduce the
Lifshitz point. For the present case, the upper critical dimension
for classical behavior is at $d=5$, and the classical location of
the Lifshitz point is at $\eta_{1}=\eta_{2}=0$. Low dimensional cases
should have enhanced fluctuations, without spontaneous magnetization
or symmetry breaking, as expected from the Mermin-Wagner theorem~\cite{Mermin-Wagner}.
This applies to Heisenberg-type models with higher-dimensional order
parameters in two dimensions. As it is valid for general, finite-range
interactions, it also holds in our case.

\section{Correlations}

All physical quantities that we wish to understand in the double adiabatic
limit treated in the previous section, come from the solution of equation
(\ref{eq:cvgl}). This allows us to calculate the expectation value
of any other observable in the vicinity of the critical point. Another
way to obtain expectation values is write the functional probability
as a solution of the master equation. Below we develop both methods.

\subsection{Stationary solution of the Fokker Planck equation}

We start with equation (\ref{eq:cvgl}) and note that it is possible
to write a functional Fokker-Planck equation for the probability density
$P(\bm{X},\tau)$, 
\begin{equation}
\frac{\partial P}{\partial\tau}(\bm{X},\tau)=\sum_{i}\frac{\delta}{\delta X_{i}}\left[\left(\left|\bm{X}\right|^{2}-\tilde{\mathcal{D}}\right)X_{i}+\frac{1}{2}\frac{\delta}{\delta X_{i}}\right]P(\bm{X},\tau)
\end{equation}
and look for the equilibrium distribution in the form $P(\bm{X})=N\exp[-{\cal H}(\bm{X})]$,
where ${\cal H}$ is a potential function. The solution for the distribution
of $\bm{X}$ is given by: 
\begin{align}
P({\bm{X}}) & =N\exp\left[-\int d^{2}x\left(\eta_{1}{\bm{X}}\cdot{\bm{X}}+\frac{1}{2}\left({\bm{X}}\cdot{\bm{X}}\right)^{2}+\right.\right.\nonumber \\
 & \left.\left.+\eta_{2}\nabla{\bm{X}}\cdot\nabla{\bm{X}}+\eta_{3}\nabla^{2}{\bm{X}}\cdot\nabla^{2}{\bm{X}}\right)\right]\,.
\end{align}
This expression is similar to the Landau-Ginzburg free energy of a
next nearest neighbor interaction in a planar magnetic interaction,
with $\bm{X}$ playing the role of a two component vector order parameter.
Owing to this parallel, the planar type-II parametric system can provide
a superb model platform for investigating fluctuations and universal
behavior in this paradigmatic system.

\subsection{Stochastic moments in the Gaussian approximation}

In order to evaluate the moments and spatial correlations we will
approximate the nonlinear terms of Eq. (\ref{eq:cvgl}) using a Gaussian
approximation together with a Green's function approach. Hence we
can write Eq. (\ref{eq:cvgl}) as follows: 
\begin{eqnarray}
\frac{\partial\left\langle X_{i}(\bm{r},\tau)X_{j}(\bm{r}^{\prime},\tau)\right\rangle }{\partial\tau} & = & \mathcal{\tilde{D}}_{r}\left\langle X_{i}(\bm{r},\tau)X_{j}(\bm{r}^{\prime},\tau)\right\rangle \nonumber \\
 & - & \left\langle X_{i}(\bm{r},\tau)\left|\bm{X}\left(\bm{r},\tau\right)\right|^{2}X_{j}(\bm{r}^{\prime},\tau)\right\rangle .\nonumber \\
\end{eqnarray}
Here $\langle\ldots\rangle$ denotes the stochastic average over the
Gaussian fluctuations and we have used the notation $\mathcal{\tilde{D}}_{r}$
in order to avoid ambiguities. We explicitly denote that the operator
$\mathcal{\tilde{D}}$ acts on the spatial coordinate $r$. In the
Gaussian approximation, fluctuations around the most probable configuration
are approximately treated as independent modes with Gaussian distributions
\cite{GaussianApproxRef}. This allows us to replace ensemble averages
by a Gaussian ansatz, in which higher order moments are approximated
by the expressions for a Gaussian distribution, e.g. $\langle X^{4}\rangle\simeq3\langle X^{2}\rangle^{2}$.
On defining $G_{ij}\equiv G_{ij}\left(\bm{r},\bm{r}^{\prime}\right)=\left\langle X_{i}(\bm{r},\tau)X_{j}(\bm{r}^{\prime},\tau)\right\rangle $
we can write the above equation, for $i=1$, as: 
\begin{eqnarray*}
\frac{\partial G_{1j}}{\partial\tau}=\mathcal{\tilde{D}}_{r}G_{1j}-G_{1j}\left(3\left\langle X_{1}^{2}\right\rangle +\left\langle X_{2}^{2}\right\rangle \right).
\end{eqnarray*}
Here we have assumed rotational symmetry of the problem in the $\left(X_{1},X_{2}\right)$
plane, so that $\left\langle X_{1}X_{2}\right\rangle =0$. Using again
the assumption of rotational symmetry in the $X_{1}-X_{2}$ plane,
we can also write the Green's function results as: 
\begin{eqnarray}
\frac{\partial G_{1j}}{\partial\tau} & = & \left[\mathcal{\tilde{D}}_{r}-2\left(\left\langle X_{1}^{2}\right\rangle +\left\langle X_{2}^{2}\right\rangle \right)\right]G_{1j}.
\end{eqnarray}
This corresponds to an equivalent stochastic equation, valid in the
steady-state for a rotationally symmetric system: 
\begin{equation}
\frac{\partial{\tilde{\bm{X}}(\bm{r},\tau)}}{\partial\tau}=\tilde{\mathcal{D}}\tilde{\bm{X}}(\bm{r},\tau)-2\left\langle \tilde{\bm{X}}\cdot\tilde{\bm{X}}\right\rangle \tilde{\bm{X}}(\bm{r},\tau)+\tilde{\bm{\zeta}}(\bm{r},\tau).
\end{equation}
We use the variable $\tilde{\bm{X}}$ to denote that we are using
the Gaussian approximation. In order to evaluate the Gaussian correlation
functions in near and far field, we will define a parameter $\eta_{1}^{\prime}=\eta_{1}+2\langle\tilde{\bm{X}}\cdot\tilde{\bm{X}}\rangle$.
Using the expression for $\mathcal{\tilde{D}}$ of Eq. (\ref{eq:tildeD})
we can write the above equation as: 
\begin{equation}
\frac{\partial\tilde{\bm{X}}(\bm{r},\tau)}{\partial\tau}=\left(-\eta'_{1}+\eta_{2}\nabla_{\bm{r}}^{2}-\eta_{3}\nabla_{\bm{r}}^{4}\right)\tilde{\bm{X}}(\bm{r},\tau)+\tilde{\bm{\zeta}}(\bm{r},\tau).\label{eq:DerX2}
\end{equation}
Next, we note that because of translational symmetry the term $\left\langle \tilde{\bm{X}}\cdot\tilde{\bm{X}}\right\rangle $
is independent of the spatial coordinate so that $\left\langle \tilde{\bm{X}}\cdot\tilde{\bm{X}}\right\rangle =\left\langle \left|\tilde{X}\right|^{2}\right\rangle $
can be calculated analytically. In order to evaluate it, we use the
Fourier transform 
\begin{equation}
\tilde{\bm{X}}(\bm{r},t)=\frac{1}{2\pi}\int e^{i\vec{k}\cdot\bm{r}}\tilde{\bm{X}}(\vec{k},t)d^{2}\vec{k},
\end{equation}
so that in momentum space we write Eq. (\ref{eq:DerX2}) as: 
\begin{equation}
\frac{\partial{\tilde{\bm{X}}(\bm{k},\tau)}}{\partial\tau}=-\left(\eta'_{1}+\eta_{2}k^{2}+\eta_{3}k^{4}\right)\tilde{\bm{X}}(\bm{k},\tau)+\tilde{\bm{\zeta}}(\bm{k},\tau).\label{eq:DerXk}
\end{equation}

\subsection{Lifshitz point}

We now consider the line of points where $\eta_{2}=0$ and $\eta_{3}=\frac{1}{2}$.
These are the points we have defined as corresponding physically to
zero detuning, with a pump in the vicinity of the $\mu=1$. In this
case the solution for $\tilde{\bm{X}}(\vec{k},t)$ is given by: 
\begin{equation}
\tilde{\bm{X}}(\bm{k},\tau)=\int_{-\infty}^{\tau}\tilde{\bm{\zeta}}(\bm{k},\tau')e^{-\left(\eta'_{1}+\frac{k^{4}}{2}\right)(\tau-\tau')}d\tau'\,.\label{eq:momentumsolution}
\end{equation}
In this way, we obtain: 
\begin{align}
\left\langle \tilde{\bm{X}}\cdot\tilde{\bm{X}}\right\rangle =\frac{1}{4\pi^{2}}\int_{0}^{\infty}\frac{2\pi kdk}{\eta'_{1}+\frac{k^{4}}{2}}\,.
\end{align}
On performing the integration there is a resulting self-consistency
condition: 
\begin{equation}
\left\langle \tilde{\bm{X}}\cdot\tilde{\bm{X}}\right\rangle =\frac{1}{4\sqrt{2(\eta_{1}+2\left\langle \tilde{\bm{X}}\cdot\tilde{\bm{X}}\right\rangle )}}.
\end{equation}
At the Lifshitz point, where $\eta_{1}=0$, we find that: 
\begin{equation}
\left\langle \tilde{\bm{X}}\cdot\tilde{\bm{X}}\right\rangle =\langle\left|\tilde{X}\right|^{2}\rangle=0.25.
\end{equation}
As explained below, this is remarkably close to accurate numerical
simulations of the correlations, including non-Gaussian fluctuations.
\begin{figure}[htbp]
\includegraphics[width=1\linewidth]{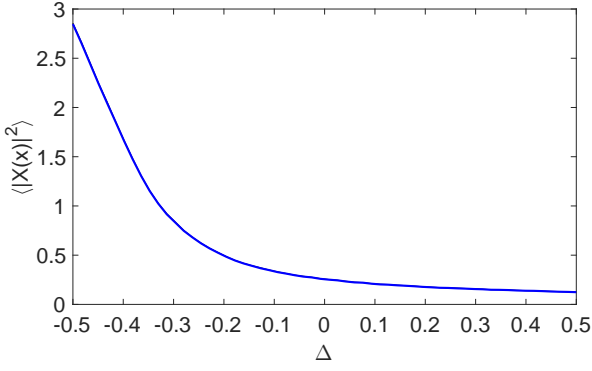} \protect\caption{Dimensionless intensity versus detuning $\Delta$. The Lifshitz point
corresponds to zero detuning. Here we have used $\mu=1$ and $g=0.01$.
Results were obtained from a simulation with $60$ samples on a $200\times50\times50$
numerical grid of $15000\times96\times96$ points. \label{fig:Detuning}}
\end{figure}

We also consider the case where $\mu=1$, $\eta_{2}\neq0$ and $\eta_{3}=\frac{1}{2}$.
This corresponds to the case of non-zero detuning $\Delta$. In Fig.
(\ref{fig:Detuning}) we plot the modulus squared of the two dimensional
order parameter $\bm{X}$ vs the detuning $\Delta$.

\begin{figure}[h!]
\includegraphics[width=1\linewidth]{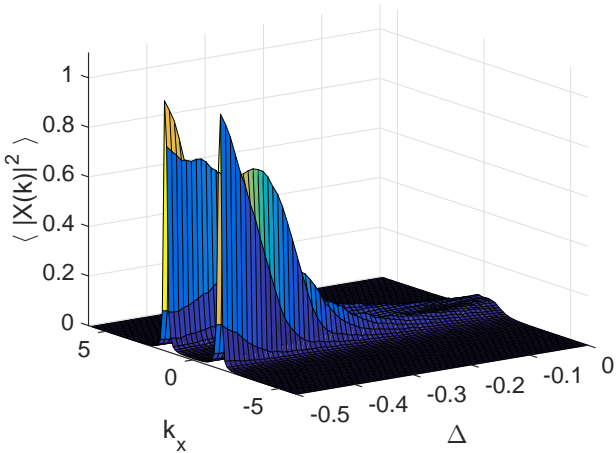} \protect\caption{Dimensionless intensity versus detuning $\Delta$ and transverse momentum
$k_{x}$. For negative detuning there is are fluctuations for lower
momentum values. Results were obtained from a simulation with $800$
samples on a $100\times50\times50$ numerical grid of $15000\times96\times96$
points. Here we have used $\mu=1$ and $g=0.01$. \label{fig:Detuningkxdelta}}
\end{figure}

\begin{figure}[htbp]
\includegraphics[width=0.6\linewidth]{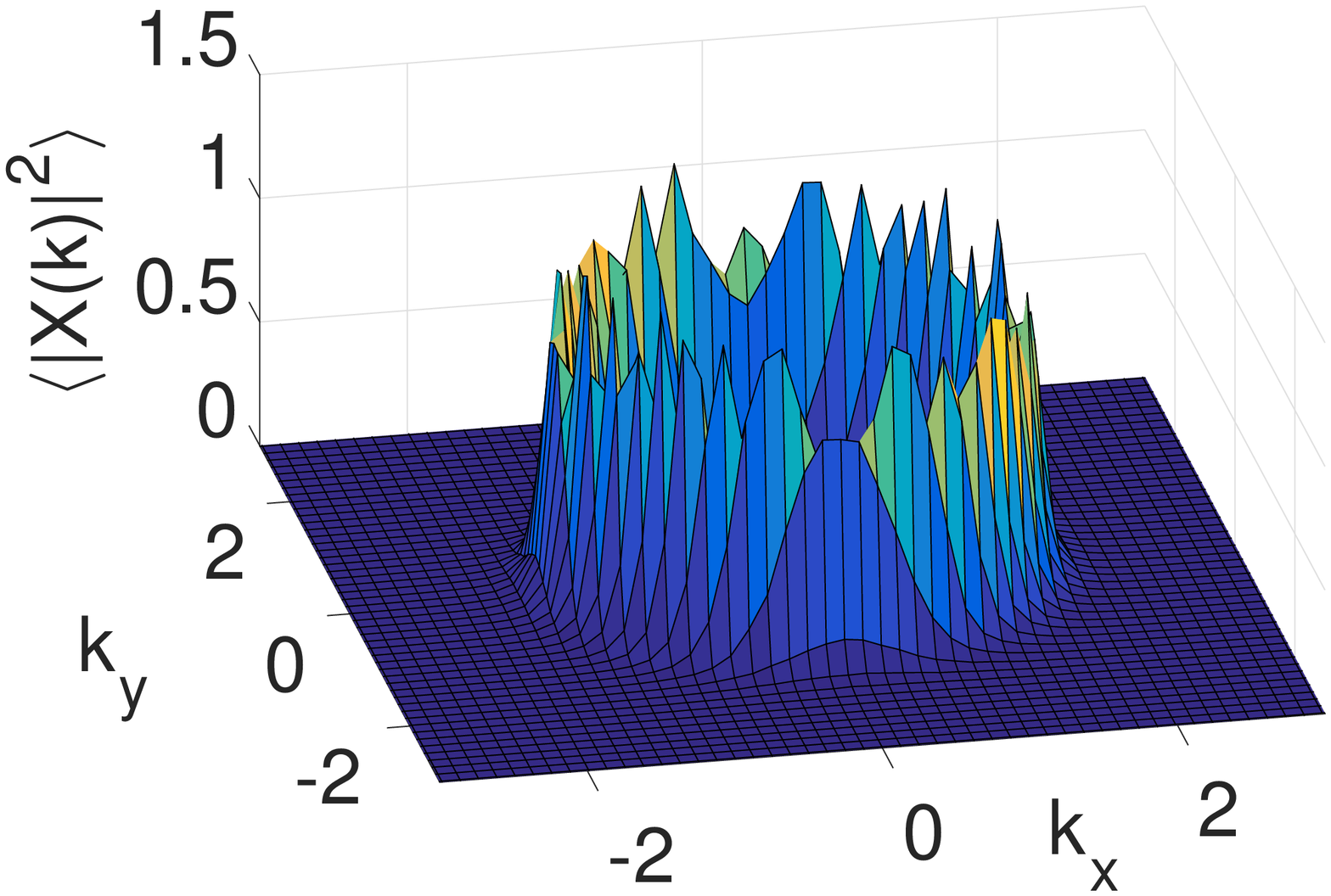}\includegraphics[width=0.4\linewidth,height=1\columnwidth,keepaspectratio]{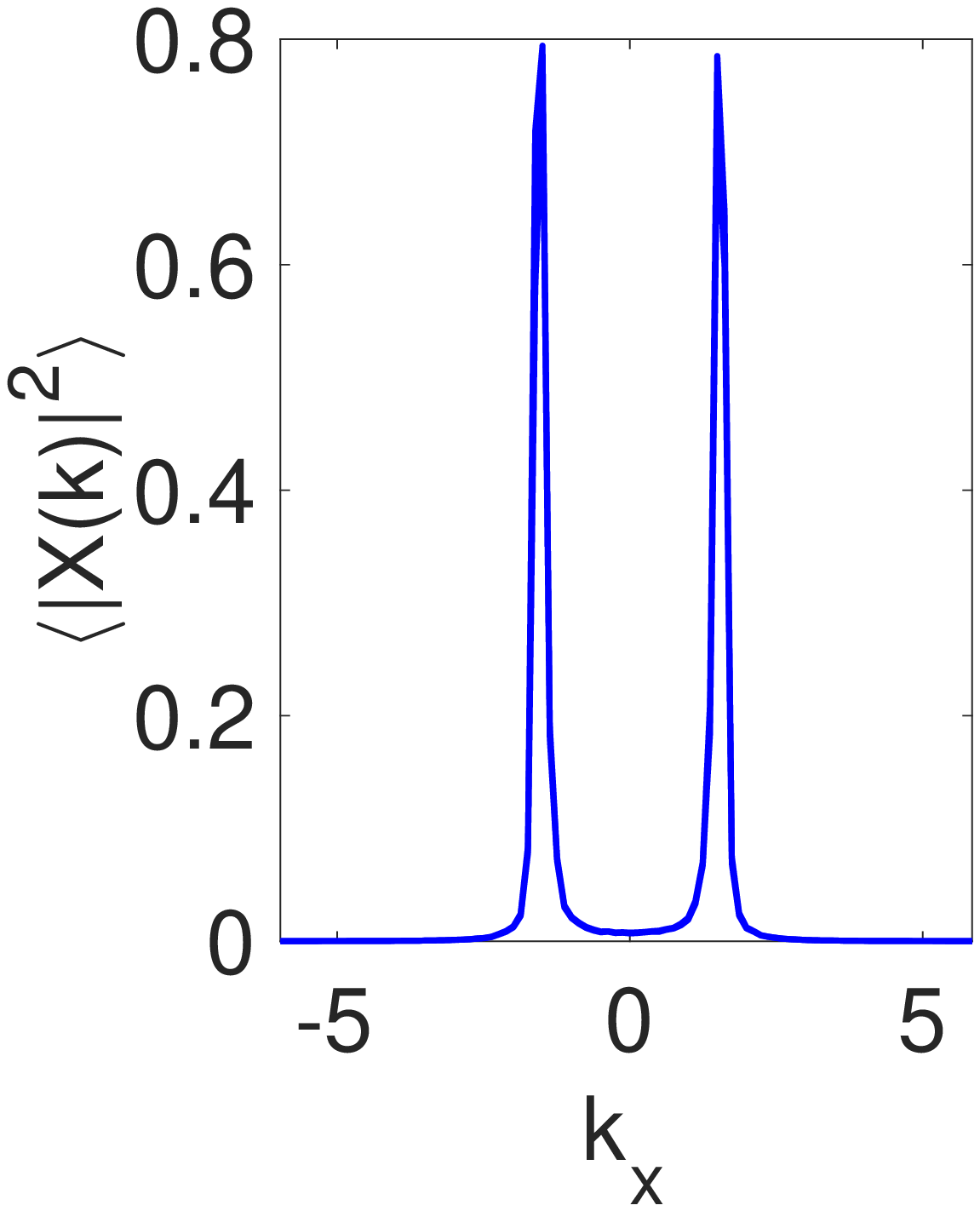}
\caption{Left: Dimensionless intensity versus transverse momentum $\mathbf{k}$
for a fixed value of the detuning $\Delta=0.5$. Here we have used
$\mu=1$ and $g=0.01$. A ring is formed for lower values of the momentum.
The fluctuations (peaks) are not symmetric due to noise. Right: Dimensionless
intensity versus transverse momentum $k_{x}$ for a fixed value of
the detuning $\Delta=0.45,$ and $k_{y}=0$. Other parameters are
as in Fig. (\ref{fig:Detuningkxdelta}). \label{fig:DetuningRing}}
\end{figure}

In order to obtain this figure, we solve the stochastic equations
given in Eq. (\ref{eq:cvgl}), and scan the detuning which is proportional
to the parameter $\eta_{2}$ defined in Eq. (\ref{eq:etas}), so that
$\Delta=-0.5+0.005t$. We notice that the fluctuations depends on
the detuning. For a positive detuning there is a decrease of the fluctuations
while for a negative detuning the fluctuations increase. A classical
Swift\textendash Hohenberg equation with a complex order parameter
and nonzero detuning has been treated \cite{StaliunasR2}, as has
the hyperbolic complex Swift\textendash Hohenberg equation\cite{StaliunasHypCSHEq}
and other related studies \cite{StaliunasExpCSHE,StaliunasBook}.
For negative detuning there is a ring with strong fluctuations, shown
in figures (\ref{fig:Detuningkxdelta}) and (\ref{fig:DetuningRing}).

\subsection{Spatial correlations in the Gaussian approximation}

From Eq (\ref{eq:momentumsolution}), the Gaussian correlation function
in the momentum space (far field), in the stationary regime is therefore
given by: 
\begin{equation}
\langle\tilde{\bm{X}}(\mathbf{k})\left[\tilde{\bm{X}}(\mathbf{k}')\right]^{*}\rangle=\frac{1}{2}\frac{\delta(\mathbf{k}+\mathbf{k}')}{\eta_{1}^{\prime}+\eta_{2}k^{2}+\eta_{3}k^{4}}.
\end{equation}
This should be readily observable experimentally, since the far-field
region of the output field from the planar cavity is simply the Fourier
transform of the internal cavity field. The correlation function for
the enhanced quadrature in configuration space, or near field, is:
\begin{equation}
\langle\tilde{\bm{X}}(\bm{r})\tilde{\bm{X}}(\bm{r}')\rangle=\frac{1}{8\pi^{2}}\int d^{2}\mathbf{k}\frac{e^{-i\mathbf{k}\cdot\left(\bm{r}-\bm{\mathbf{r}}'\right)}}{\eta_{1}^{\prime}+\eta_{2}k^{2}+\eta_{3}k^{4}}.
\end{equation}
On integrating over the angle variable we obtain: 
\begin{equation}
\langle\tilde{\bm{X}}(\bm{r})\tilde{\bm{X}}(\bm{r}')\rangle=\frac{1}{4\pi}\int_{0}^{\infty}dk\frac{kJ_{0}\left(k|\bm{r}-\bm{r}'|\right)}{\eta_{1}^{\prime}+\eta_{2}k^{2}+\eta_{3}k^{4}},
\end{equation}
where $J_{0}$ is the Bessel function of zero order. The result of
the above integration is, in the limit $\eta_{2}=0$: 
\begin{equation}
\langle\tilde{\bm{X}}(\bm{r})\tilde{\bm{X}}(\bm{r}')\rangle=-\sqrt{\frac{\eta_{3}}{\eta_{1}^{\prime}}}\frac{1}{4\pi\eta_{3}}\;kei\left(\left(\frac{\eta_{1}^{\prime}}{\eta_{3}}\right)^{1/4}|\bm{r}-\bm{r}'|\right)\,,
\end{equation}
where $kei(x)$ is Thomson's function \cite{Gradshtein}. An interesting
remark is that as $\eta_{1}^{\prime}\rightarrow0$, the spatial correlation
decays with a power law: 
\begin{equation}
\langle\tilde{\bm{X}}(\bm{r})\tilde{\bm{X}}(\bm{r}')\rangle\propto|\bm{r}-\bm{r}'|^{-0.5}\,.
\end{equation}
We note, however, that this system is predicted to have an upper critical
dimension \cite{Hornreich} with mean-field critical exponents at
spatial dimension $d>5$. Since the system is well below this critical
dimension, one may expect non-Gaussian behavior that is not predicted
by the approximations used in this section.

\subsection{Non-Gaussian behavior and universality}

In order to verify these analytic results and investigate non-Gaussian
correlations, numerical simulations were carried out of the original
stochastic partial differential equations of Eq. (\ref{eq:cvgl}).
Small time steps are needed to treat the quartic growth of the squared
Laplacian term in momentum, together with large sample numbers to
obtain a low sampling error. In initial investigations, we used a
$10\times10\times10$ numerical grid of $8000\times64\times64$ points
in $t,x,y$ respectively. Employing a fine numerical grid of $16000\times64\times64$
points to check convergence in time-step, the final steady-state correlation
result converged to $\langle\bm{X}\cdot\bm{X}\rangle=0.264\pm0.005$.
This was close to the Gaussian value, but with a relatively large
sampling error.

In order to understand the quantitative difference between the exact
and Gaussian results, a more precise differencing technique was used.
This variance reduction or differencing technique simulates the difference
between the full sample path and the Gaussian approximation \cite{CDD}.
The results were in agreement with a direct simulation, but gave much
more rapid convergence. This was carried out as follows. First a mean
field variable, $\tilde{X}$ was simulated, in the Gaussian approximation,
using: 
\begin{equation}
\frac{\partial\tilde{X}}{\partial\tau}=\mathcal{\tilde{D}}\tilde{X}-2\tilde{X}\left\langle \left|\tilde{X}\right|^{2}\right\rangle -2\tilde{X}^{*}\left\langle \tilde{X}^{2}\right\rangle +\zeta_{+}\,,
\end{equation}
where the averages were carried out both over a spatial numerical
grid and a set of parallel trajectories. From the analytic calculations
above, this should converge precisely to the analytic result of $\left\langle |\tilde{X}|^{2}\right\rangle =0.25$
, which was confirmed to a numerical accuracy of $\pm5\times10^{-3}$.
Here we have considered that $\eta_{1}=\eta_{2}=0$ and $\eta_{3}=\frac{1}{2}$,
corresponding to the classical Lifshitz point. 
\begin{figure}[h!]
\includegraphics[width=1\linewidth]{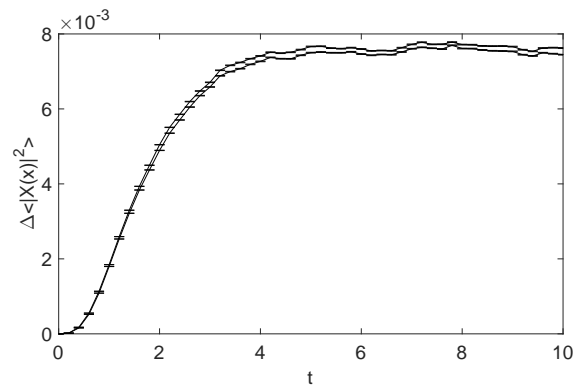} \protect\caption{Growth of non-Gaussian correlations, $\langle\left|X\right|^{2}-|\tilde{X}|^{2}\rangle$
versus time $\tau$, starting from $\bm{X}=0$. Error bars were obtained
from comparing a coarse ($5000$ step) and fine ($10000$ step) simulation.
The sampling error is indicated by the upper and lower solid lines,
with a standard deviation of $\pm0.00025$. This gives the steady-state
correlation result $\langle\bm{X}\cdot\bm{X}\rangle=\langle\left|X\right|^{2}\rangle=0.2574\pm0.0003$.
Results were obtained from a simulation with $3200$ samples on a
$10\times20\times20$ fine numerical grid of $10000\times48\times48$
points. \label{fig:NonGaussianCorr}}
\end{figure}

\begin{figure}[h!]
\includegraphics[width=0.5\linewidth]{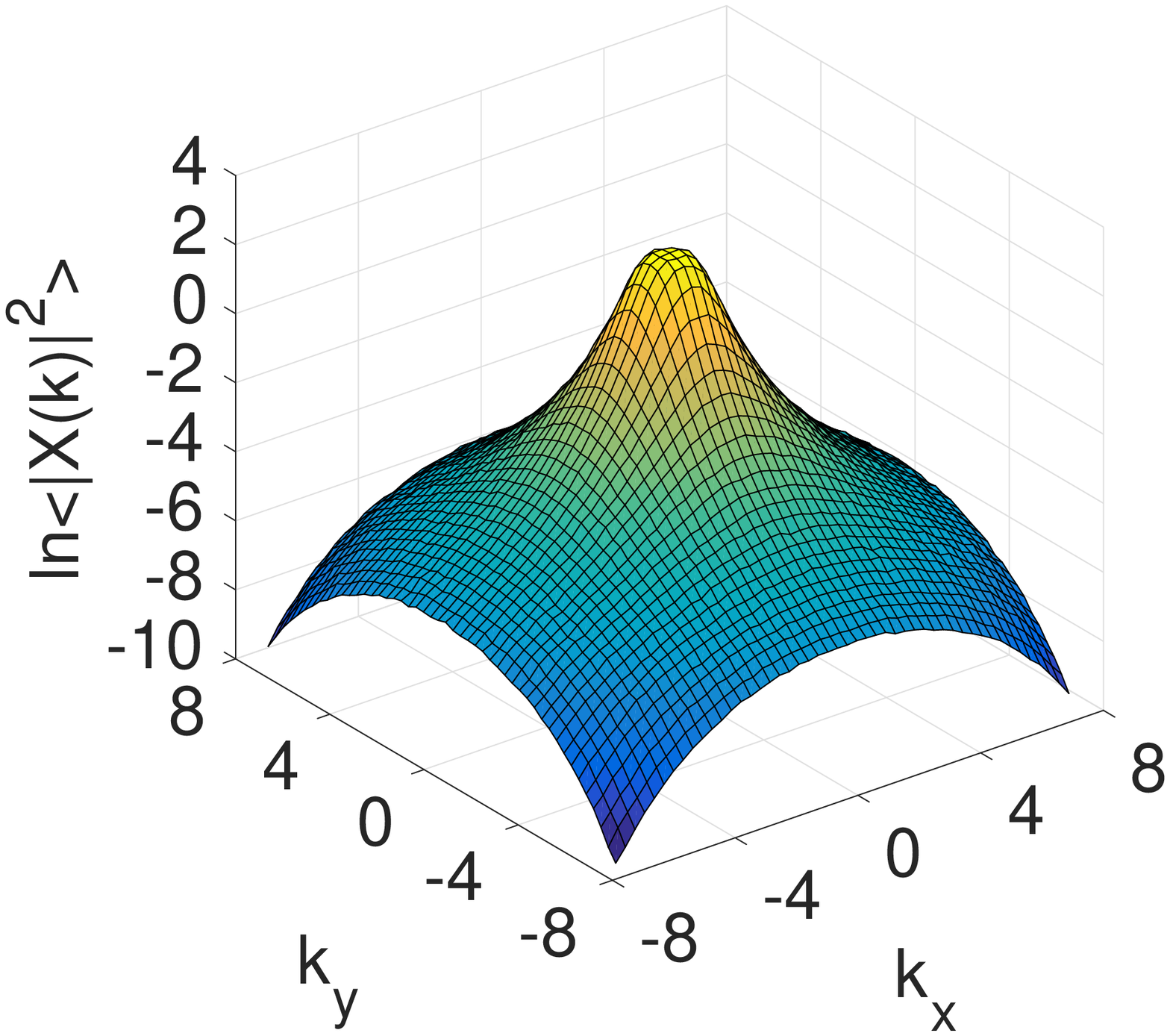}\includegraphics[width=0.5\linewidth]{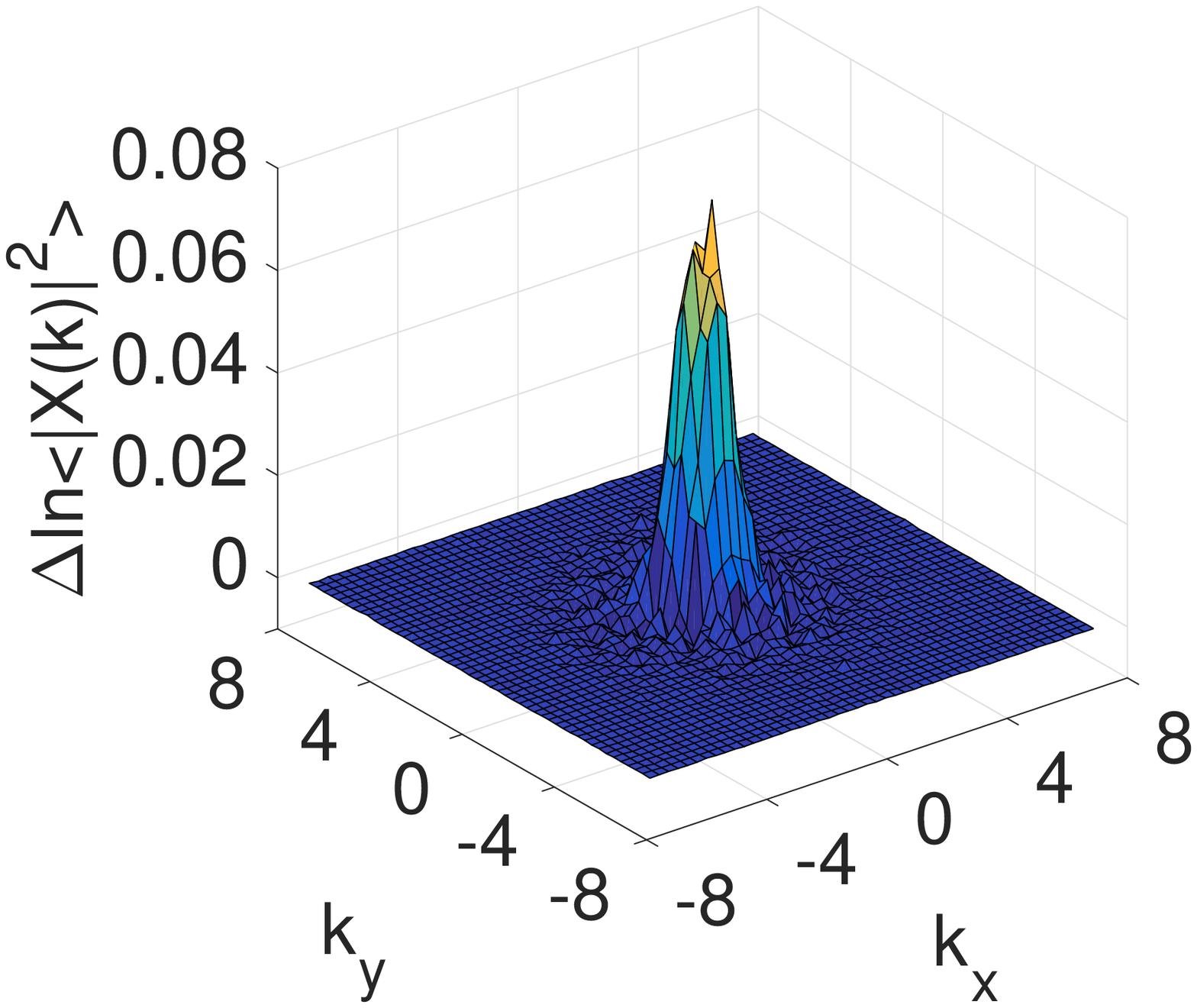}
\caption{Left: Steady-state momentum correlations, $\langle\left|X(\mathbf{k})\right|^{2}\rangle$
versus transverse momentum $\mbox{\textbf{k}}$, starting from $X=0$.
Right: Steady-state non-Gaussian correlations in momentum space, $\Delta\ln\left\langle \left|X(\mathbf{k})\right|^{2}\right\rangle =\ln\left\langle \left|X(\mathbf{k})\right|^{2}\right\rangle -\ln\left\langle |\tilde{X}(\mathbf{k})|^{2}\right\rangle $
versus momentum $\mathbf{k}$. Here we consider $\eta_{1}=\eta_{2}=0$
and $\eta_{3}=\frac{1}{2}$. Other parameters are as in Fig. (\ref{fig:NonGaussianCorr})
\label{fig:MomentumCorr}}
\end{figure}

Next, the full variable $X$ was simulated. This was achieved by introducing
a difference variable defined as $\Delta_{X}=X-\tilde{X}$, which
has the stochastic equation: 
\begin{equation}
\frac{\partial\Delta_{X}}{\partial\tau}=\mathcal{\tilde{D}}X-X\left|X\right|^{2}+\zeta_{+}-\frac{\partial\tilde{X}}{\partial\tau}\,.
\end{equation}
By using an identical noise source to those in the equations for the
mean-field $\tilde{X}$, the difference simulation permits a more
precise calculation with reduced variance. The result, shown in Figs.
(\ref{fig:NonGaussianCorr}) and (\ref{fig:MomentumCorr}) was that,
at the critical point, $\langle\bm{X}\cdot\bm{X}\rangle=0.2574\pm0.0003$.
This result, of much greater accuracy, only required $3200$ samples
with a $10\times20\times20$ numerical grid of $10000\times48\times48$
points. The discretization error was estimated from using several
grids with different transverse sizes and time-steps.

In summary, the full statistical calculation gives increased critical
fluctuations due to non-Gaussian effects, but this increase is relatively
small. At large transverse momentum there is no measurable difference
between the Gaussian and exact results, shown in Fig. (\ref{fig:MomentumCorr}).
The deviation from the Gaussian approximation vanishes rapidly as
higher order transverse momenta are investigated. Small values of
this difference are observed only at zero transverse direction.

\section{Conclusion}

We have shown that parametric down-conversion in a type-II parametric
planar cavity leads to a Swift and Hohenberg type of stochastic equation
for the leading terms in the critical fluctuations, but with a vector
order parameter. This combines the rotationally invariant symmetry
properties of the X-Y model with the higher-order Laplacian of a Lifshitz
magnetic phase transition. Surprisingly, these fluctuations are not
thermal in origin, but come instead from the quantum fluctuations
associated with parametric amplification.

This model can be approximately treated for the critical fluctuations
with a Gaussian factorization. However, a careful numerical treatment
shows that non-Gaussian critical fluctuations occur. These are responsible
for enhanced intensity correlations, but are reduced at large transverse
momenta due to the momentum dependence of the linear propagator. As
techniques improve, we expect that this novel, non-equilibrium critical
point will become accessible to experimental studies.

\appendix

\section*{Appendix A: Quantum Langevin form}

\setcounter{equation}{0} \global\long\def\theequation{A{\arabic{equation}}}

In the quantum Langevin form the corresponding operator equations
of the system would be: 
\begin{align}
\frac{\partial\hat{A}_{0}}{\partial t} & =-\tilde{\gamma}_{0}\hat{A}_{0}+{\mathcal{E}}(\bm{x})-\chi^{*}\hat{A}_{1}\hat{A}_{2}+\frac{iv_{0}^{2}}{2\omega_{0}}\nabla^{2}\hat{A}_{0}+\sqrt{2\gamma_{0}}\hat{A}_{0}^{in}\,,\nonumber \\
\frac{\partial\hat{A}_{1}}{\partial t} & =-\tilde{\gamma}_{1}\hat{A}_{1}+\chi\hat{A}_{0}\hat{A}_{2}^{\dagger}+\frac{iv_{1}^{2}}{2\omega_{1}}\nabla^{2}\hat{A}_{1}+\sqrt{2\gamma_{1}}\hat{A}_{1}^{in}\,,\nonumber \\
\frac{\partial\hat{A}_{2}}{\partial t} & =-\tilde{\gamma}_{2}\hat{A}_{2}+\chi\hat{A}_{0}\hat{A}_{1}^{\dagger}+\frac{iv_{2}^{2}}{2\omega_{2}}\nabla^{2}\hat{A}_{2}+\sqrt{2\gamma_{2}}\hat{A}_{2}^{in}\,.
\end{align}
Here we use a rotating frame such that the three field operators are
treated as in a frame rotating with frequency $\omega_{i}^{0}$. The
relative detuning between the pump laser at $2\omega_{L}$ and the
intracavity pumped mode $\omega_{0}$ is $\Delta_{0}=\left(\omega_{0}-2\omega_{L}\right)/\gamma_{0}$,
and the down-converted modes have relative detunings $\Delta_{i}=\left(\omega_{i}-\omega_{i}^{0}\right)/\gamma_{i}$.
The terms $\tilde{\gamma}_{i}=\gamma_{i}\left(1+i\Delta_{i}\right)$
represent the complex cavity decay for each mode, including detunings.
However, the quantum Langevin approach has the drawback that it deals
with operator equations that are intractable analytically. It was
shown in section \ref{sec:IIIPositiveP} that the phase-space that
the phase-space representation method generates similar equations,
but with a more useful c-number form.

The input-output relations that describe the external modes outside
the cavity are~\cite{InputOutputTheo,YurkeIO,GardinerQNoise,InputOutputBooks,PeterBook}:
$\hat{A}_{i}^{out}=\sqrt{2\gamma_{i}}\hat{A}_{i}-\hat{A}_{i}^{in},$
where $\hat{A}_{i}^{in}$ and $\hat{A}_{i}^{out}$ are the corresponding
input and output fields, with input correlations: 
\begin{align}
\left\langle \hat{A}_{i}^{in}(\bm{x},t)\hat{A}_{j}^{in\dagger}(\bm{x}',t)\right\rangle  & =\left(\bar{n}_{i}^{th}+1\right)\delta_{ij}\delta\left(\bm{x}-\bm{x}'\right),\nonumber \\
\left\langle \hat{A}_{i}^{in\dagger}(\bm{x},t)\hat{A}_{j}^{in}(\bm{x}',t)\right\rangle  & =\bar{n}_{i}^{th}\delta_{ij}\delta\left(\bm{x}-\bm{x}'\right).
\end{align}
In our calculations we assume that the reservoirs are in the vacuum
state. However, non-zero reservoir temperatures can be readily included.
While we do not use these input-output equations here, we note that
they are important when dealing with external measurements.

\section*{Appendix B: Positive-P representation}

\setcounter{equation}{0} \global\long\def\theequation{C{\arabic{equation}}}

The positive P-representation generates a genuine (second order) Fokker-Planck
equation with positive-definite diffusion, provided the distribution
vanishes sufficiently rapidly at the phase-space boundaries. This
can then be mapped into a set of c-number Langevin equations similar
to the quantum Heisenberg equations, except for additional stochastic
terms.

This approach uses a multi-mode coherent state $\left|\tilde{\bm{\alpha}}_{0},\tilde{\bm{\alpha}}_{1},\tilde{\bm{\alpha}}_{2}\right\rangle \equiv\left|\tilde{\bm{\alpha}}\right\rangle $,
defined as an eigenstate of the annihilation operators $\hat{a}_{i}\left(\bm{k}\right)$,
where $\tilde{\alpha}_{i}\equiv\tilde{\alpha}_{i}\left(\bm{k}\right)$
so that the following eigenvalue equation is obtained: 
\begin{equation}
\hat{a}_{i}\left(\bm{k}\right)\left|\tilde{\bm{\alpha}}\right\rangle =\tilde{\alpha}_{i}\left(\bm{k}\right)\left|\tilde{\bm{\alpha}}\right\rangle \,.
\end{equation}
The positive-P representation then expands the density matrix in terms
of coherent state projection operators~\cite{PosP}, which is always
possible as a positive distribution: 
\begin{equation}
\hat{\rho}=\int d^{6M}\tilde{\bm{\alpha}}d^{6M}\tilde{\bm{\alpha}}^{+}\frac{\left|\tilde{\bm{\alpha}}\right\rangle \left\langle \tilde{\bm{\alpha}}^{+*}\right|}{\left\langle \tilde{\bm{\alpha}}^{+*}\right.\left|\tilde{\bm{\alpha}}\right\rangle }P\left(\tilde{\bm{\alpha}},\tilde{\bm{\alpha}}^{+}\right)\,.
\end{equation}
In the positive P-representation the coherent amplitudes $\tilde{\alpha}_{i}\left(\bm{k},t\right)$
satisfy time-dependent stochastic differential equations. It is simplest
to write these equations in a form analogous to classical equations
by introducing stochastic fields $A_{i}$ , $A_{i}^{+}$ defined as:
\begin{align}
A_{i}(\bm{x},t)=\frac{1}{L}\sum_{\bm{k}}e^{i\bm{k}\cdot\bm{x}}\tilde{\alpha}_{i}(\bm{k},t)\,,
\end{align}
together with a stochastic conjugate $A_{i}^{+}$ which is a c-number,
rather than an operator field. It is only conjugate to $A_{i}$ in
the mean: it is stochastically equivalent to the conjugate operator.

\subsection{Noises of the stochastic equations}

We note that while our derivation of the set of stochastic equations
given in Eq. (\ref{eq:+P-equations}) is formally based on the Itô
stochastic calculus, in this case either Itô or Stratonovich stochastic
calculus gives identical results. These complex noise terms can be
constructed from four delta-correlated real Gaussian noise fields
$\left(\xi_{x},\xi_{y},\xi_{x}^{+},\xi_{y}^{+}\right)$, with the
mapping: 
\begin{align}
\xi_{1,2}(\bm{x},t) & =\left[\xi_{x}(\bm{x},t)\pm i\xi_{y}(\bm{x},t)\right]/\sqrt{2}\,,\nonumber \\
\xi_{1,2}^{+}(\bm{x},t) & =\left[\xi_{x}^{+}(\bm{x},t)\pm i\xi_{y}^{+}(\bm{x},t)\right]/\sqrt{2}\,.
\end{align}
It follows that the stochastic fields in the positive P-representation
for the $\xi_{1}$ and $\xi_{2}$ fields are complex conjugate, i.e.,
\begin{equation}
\xi_{1}(\bm{x},t)=\xi_{2}^{*}(\bm{x},t)\,,\qquad{\rm and}\qquad\xi_{1}^{+}(\bm{x},t)=\left(\xi_{2}^{+}(\bm{x},t)\right)^{*}.\label{eq:conjugate-noise}
\end{equation}
The physics of the noise is that it describes a departure from coherent
behavior. The deterministic terms correspond to the evolution of coherent
states. However, the true quantum state does not remain coherent.
Instead it develops, via the noise terms, into squeezed, entangled,
and even more complex states. Despite this complexity, there are simple,
universal properties caused by these noise terms at the critical point.

We note that there are no `normal' noise correlations, that is, $\left\langle \xi_{i}(\bm{x},t)\xi_{k}^{+}(\bm{x}',t')\right\rangle =0\,$.
This is due to our assumption that the optical reservoirs are at zero
temperature, which is an excellent approximation at optical frequencies.
If there are thermal reservoirs, as can occur in microwave devices,
then additional reservoir correlations must be included, which are
proportional to the thermal occupation number. More generally, our
model includes only the minimal noise due to fundamental quantum effects,
but there can be a range of additional technical noise sources in
practical devices, caused by temperature fluctuations, laser intensity
fluctuations and laser phase noise~\cite{ReidCorrelations,PM_PRA37}.


\begin{thebibliography}{10}
\bibitem{bowman} C. Bolman and A. C. Newell, ``Natural patterns
and wavelets,'' Rev. Mod. Phys. \textbf{70}, 289 (1998); J. P. Gollub
and J. S. Langer, ``Pattern formation in nonequilibrium physics,''
Rev. Mod. Phys. \textbf{71}, S396 (1999).

\bibitem{Pattern formation}M. C. Cross and P. C. Hohenberg, ``Pattern
formation outside of equilibrium,'' Rev. Mod. Phys. \textbf{65},
851 (1993); P. C. Hohenberg and B. I. Halperin, ``Theory of dynamic
critical phenomena,'' Rev. Mod. Phys. \textbf{49}, 435 (1977).

\bibitem{SHeq} J. Swift and P. C. Hohenberg, ``Hydrodynamic fluctuations
at the convective instability,'' Phys. Rev. A \textbf{15}, 319 (1977).

\bibitem{Graham-Haken}R. Graham and H. Haken, ``Laserlight -First
example of a second order phase transition far from thermal equilibrium,''
Z. Physik \textbf{237}, 31 (1970); V. deGiorgio and M. O. Scully,
``Analogy between the Laser Threshold Region and a Second-Order Phase
Transition,'' Phys. Rev. A \textbf{2}, 1170 (1970).

\bibitem{ReidEPR}M. D. Reid, ``Demonstration of the Einstein-Podolsky-Rosen
paradox using nondegenerate parametric amplification,'' Phys. Rev.
A \textbf{40}, 913 (1989); M. D. Reid and P. D. Drummond, ``Quantum
Correlations of Phase in Nondegenerate Parametric Oscillation,''
Phys. Rev. Lett. \textbf{60}, 2731 (1988); M. D. Reid, P. D. Drummond,
W. P. Bowen, E. G. Cavalcanti, P. K. Lam, H. A. Bachor, U. L. Andersen
and G. Leuchs, ``The Einstein-Podolsky-Rosen paradox: From concepts
to applications,'' Rev. Mod. Phys. \textbf{81}, 1727 (2009).

\bibitem{Ou}Z. Y. Ou, S. F. Pereira, H. J. Kimble and K. C. Peng,
``Realization of the Einstein-Podolsky-Rosen paradox for continuous
variables ,'' Phys. Rev. Lett. \textbf{68}, 3663 (1992).

\bibitem{ReidCorrelations}M. D. Reid and P. D. Drummond, ``Correlations
in nondegenerate parametric oscillation: Squeezing in the presence
of phase diffusion,'' Phys. Rev. A \textbf{40}, 4493 (1989).

\bibitem{PM_PRA41}P. D. Drummond and M. D. Reid, ``Correlations
in nondegenerate parametric oscillation. II. Below threshold results,''
Phys. Rev. A \textbf{41}, 3930 (1990).

\bibitem{InputOutputTheo} M. J. Collett and C. W. Gardiner ``Squeezing
of intracavity and traveling-wave light fields produced in parametric
amplification'', Phys. Rev. A \textbf{30}, 1386 (1984); C. W. Gardiner
and M. J. Collett, ``Input and output in damped quantum systems:
Quantum stochastic differential equations and the master equation,''
Phys. Rev. A \textbf{31}, 3761 (1985).

\bibitem{YurkeIO} B. Yurke, ``Squeezed-coherent-state generation
via four-wave mixers and detection via homodyne detectors,'' Phys.
Rev. A \textbf{32}, 300 (1985).

\bibitem{Wu}L. A. Wu, H. J. Kimble, J. L. Hall and H. Wu, ``Generation
of Squeezed States by Parametric Down Conversion,'' Phys. Rev. Lett.
\textbf{57}, 2520 (1986).

\bibitem{Feng2003} S. Feng and O. Pfister ``Stable nondegenerate
optical parametric oscillation at degenerate frequencies in Na:KTP,'',
J. Opt. B \textbf{5,} 262\textendash 267 (2003).

\bibitem{Villar2007}A. S. Villar, K. N. Cassemiro, K. Dechoum, A.
Z. Khoury, M. Martinelli and P. Nussenzveig, ``Entanglement in the
above-threshold optical parametric oscillator ,'' J. Opt. Soc. Am.
B \textbf{24}, 249 (2007).

\bibitem{LauratHal}J. Laurat, T. Coudreau and C. Fabre, in \emph{Quantum
Information with continuous variables of atoms and light}, edited
by Ed. N. J. Cerf, G. Leuchs and E. S. Polzik (World Scientific Publishing
2007).

\bibitem{LauratFabre}J. Laurat, T. Coudreau, G. Keller, N. Treps
and C. Fabre, ``Effects of mode coupling on the generation of quadrature
Einstein-Podolsky-Rosen entanglement in a type-II optical parametric
oscillator below threshold ,'' Phys. Rev. A \textbf{71}, 022313 (2005).

\bibitem{OPOII} J. Laurat, L. Longchambon, C. Fabre and T. Coudreau,
``Experimental investigation of amplitude and phase quantum correlations
in a type II optical parametric oscillator above threshold: from nondegenerate
to degenerate operation,'' Optics Letters \textbf{30}, 1177 (2005).

\bibitem{Keller}G. Keller, V. D'Auria, N. Treps, T. Coudreau, J.
Laurat and C. Fabre, ``Experimental demonstration of frequency-degenerate
bright EPR beams with a self-phase-locked OPO,'' Optics Express \textbf{16},
9351 (2008).

\bibitem{D'Auria}V. D'Auria, A. Chiummo, M. De Laurentis, A. Porzio,
S. Solimeno and M. G. A. Paris, ``Tomographic characterization of
OPO sources close to threshold,'' Optics Express \textbf{13}, 948
(2005); V. D'Auria, C. de Lisio, A. Porzio, S. Solimeno, J. Anwar,
and M. G. A. Paris, ``Non-Gaussian states produced by close-to-threshold
optical parametric oscillators: Role of classical and quantum fluctuations,''
Phys. Rev. A \textbf{81}, 033846 (2010).

\bibitem{DegOPO}G.-L. Oppo, M. Brambilla and L. A. Lugiato, ``Formation
and evolution of roll patterns in optical parametric oscillators,''
Phys. Rev. A \textbf{49}, 2028 (1994).

\bibitem{Lifshitz}E. M. Lifshitz and L. P. Pitaevskii, \emph{Physical
Kinetics}, Volume 10 of Course of Theoretical Physics, (Pergamon Press,
Oxford 1981).

\bibitem{Michelson}A. Michelson, ``Phase diagrams near the Lifshitz
point. I. Uniaxial magnetization,'' Phys. Rev. B \textbf{16}, 577
(1977).

\bibitem{Hornreich79} R. M. Hornreich, R. Liebmann, H.G. Schuster
and W. Selke, ``Lifshitz points in ising systems,'' Z. Phys. B\textbf{
35,} 91(1979); R. M. Hornreich, ``The Lifshitz point: Phase diagrams
and critical behavior,'' Journal of Magnetic Materials \textbf{15,}
387 (1980).

\bibitem{Planar}P. D. Drummond and K. Dechoum, ``Universality of
Quantum Critical Dynamics in a Planar Optical Parametric Oscillator,''
Phys. Rev. Lett. \textbf{95}, 083601 (2005).

\bibitem{MorcilloSH}V. J. Sánchez-Morcillo, E. Roldán, G. J. de Valcárcel
and K. Staliunas, ``Generalized complex Swift-Hohenberg equation
for optical parametric oscillators,'' Phys. Rev. A \textbf{56}, 3237
(1997).

\bibitem{Fabre}M. Vaupel, A. Maitre and C. Fabre, ``Observation
of Pattern Formation in Optical Parametric Oscillators,'' Phys. Rev.
Lett. \textbf{83}, 5278 (1999); M. Martinelli, N. Treps, S. Ducci,
S. Gigan, A. Maitre and C. Fabre, ``Experimental study of the spatial
distribution of quantum correlations in a confocal optical parametric
oscillator,'' Phys. Rev. A \textbf{67}, 023808 (2003).

\bibitem{Ducci} S. Ducci, N. Treps, A. Ma\^{i}tre and C. Fabre, ``Pattern
formation in optical parametric oscillators,'' Phys. Rev. A \textbf{64},
023803 (2001).

\bibitem{LegaLifshitzPNonOpt}J. Lega, J. V. Moloney and A. C. Newell,
``Swift-Hohenberg Equation for Lasers,'' Phys. Rev. Lett. \textbf{73},
2978 (1994).

\bibitem{Hornreich}R. M. Hornreich, M. Luban and S. Shtrikman, ``Critical
Behavior at the Onset of $\stackrel{\ensuremath{\rightarrow}}{\mathrm{k}}$-Space
Instability on the $\ensuremath{\lambda}$ Line'' Phys. Rev. Lett.
\textbf{35}, 1678 (1975).

\bibitem{TP} E. K. Riedel and F. J. Wegner, ``Tricritical exponents
and scaling fields'' Phys. Rev. Lett. \textbf{29}, 349 (1972); P.
M. Chaikin and T. C. Lubensky, P\emph{rinciples of Condensed Matter
Physics} (Cambridge University Press 1995); A. Bonanno and D. Zappalà,
Nuclear Physics B \textbf{893}, 501 (2015).

\bibitem{Staliunas}K. Staliunas, ``Laser Ginzburg-Landau equation
and laser hydrodynamics,'' Phys. Rev. A \textbf{48}, 1573 (1993).
P. C. Hohenberg and A. P. Krekhov, ``An introduction to the Ginzburg\textendash Landau
theory of phase transitions and nonequilibrium patterns,'' Physics
Reports \textbf{572}, 1 (2015).

\bibitem{Lugiato}A. Gatti and L. Lugiato, ``Quantum images and critical
fluctuations in the optical parametric oscillator below threshold,''
Phys. Rev. A \textbf{52}, 1675 (1995); A. Gatti, H. Wiedemann, L.
A. Lugiato, I. Marzoli, G.-L. Oppo and S. M. Barnett, ``Langevin
treatment of quantum fluctuations and optical patterns in optical
parametric oscillators below threshold,'' Phys. Rev. A \textbf{56},
877 (1997); A. Gatti, L.A. Lugiato, G.-L. Oppo, R. Martin, P. Di Trapani
and A. Berzanskis, ``From quantum to classical images,'' Opt. Express
\textbf{1}, 21 (1997); L. A. Lugiato, A. Gatti and E. Brambilla, ``Quantum
imaging,'' J. Opt. B \textbf{4}, S176 (2002).

\bibitem{Peter}P. D. Drummond, K. J. McNeil and D. F. Walls, ``Non-equilibrium
Transitions in Sub/Second Harmonic Generation,'' Optica Acta \textbf{27},
321 (1980); P. D. Drummond, K. J. McNeil and D. F. Walls, ``Non-equilibrium
Transitions in Sub/second Harmonic Generation,'' Optica Acta \textbf{28},
211 (1981).

\bibitem{Marte}M. A. M. Marte, H. Ritsch, K. Petsas, A. Gatti, L.
Lugiato, C. Fabre and D. Leduc, ``Spatial patterns in optical parametric
oscillators with spherical mirrors: classical and quantum effects,''
Optics Express \textbf{3}, 71 (1998).

\bibitem{Pointer}N. Treps, N. Grosse, W. P. Bowen, C. Fabre, H. A.
Bachor and P. K. Lam, ``A Quantum Laser Pointer,'' Science \textbf{301},
940 (2003).

\bibitem{Swain}C. J. Mertens, T. A. B. Kennedy and S. Swain, ``Many
body theory of quantum noise,'' Phys. Rev. Lett. \textbf{71}, 2014
(1993).

\bibitem{Berezinskii}V. L. Berezinskii, ``Destruction of Long-range
Order in One-dimensional and Two-dimensional Systems Possessing a
Continuous Symmetry Group. II. Quantum Systems,'' Sov. Phys. JETP
\textbf{34}, 610 (1972).

\bibitem{K-T} J. M. Kosterlitz and D. J. Thouless, ``Ordering, metastability
and phase transitions in two-dimensional systems,'' J. Phys. C \textbf{6},
1181, (1973).

\bibitem{CDD}S. Chaturvedi, K. Dechoum and P. D. Drummond, ``Limits
to squeezing in the degenerate optical parametric oscillator,'' Phys.
Rev. A \textbf{65}, 033805 (2002); K. Dechoum, P. D. Drummond, S.
Chaturvedi and M. D. Reid, ``Critical fluctuations and entanglement
in the nondegenerate parametric oscillator,'' Phys. Rev. A \textbf{70},
053807 (2004).

\bibitem{Mermin-Wagner} N. D. Mermin and H. Wagner, ``Absence of
Ferromagnetism or Antiferromagnetism in One- or Two-Dimensional Isotropic
Heisenberg Models,'' Phys. Rev. Lett. \textbf{17}, 1133 (1966).

\bibitem{PosP}P. D. Drummond and C. W. Gardiner, ``Generalised P-representations
in quantum optics,'' J. Phys. A \textbf{13}, 2353 (1980).

\bibitem{Haken} H. Haken, \emph{Laser Theory}, (Springer Berlin Heidelberg
1984).

\bibitem{Ritsch}C. Lamprecht, M. K. Olsen, P. D. Drummond and H.
Ritsch, ``Positive-P and Wigner representations for quantum-optical
systems with nonorthogonal modes,'' Phys. Rev. A \textbf{65}, 053813
(2002).

\bibitem{GardinerQNoise}C. W. Gardiner and P. Zoller, \emph{Quantum
Noise} (Springer, Berlin 2000).

\bibitem{InputOutputBooks} B. Yurke in \emph{Quantum Squeezing},
edited by P. D. Drummond and Z. Ficek (Springer, Berlin, 2004); D.
F. Walls G. J Milburn, \emph{Quantum Optics} (Springer, Berlin, 2008)

\bibitem{PeterBook}P. D. Drummond and M. Hillery, \emph{The Quantum
Theory of Nonlinear Optics} (Cambridge University Press 2014).

\bibitem{CarusottoRMP}I. Carusotto and C. Ciuti, ``Quantum fluids
of light,'' Rev. Mod. Phys. \textbf{85}, 299 (2013).

\bibitem{Hillery}M. Hillery and L. D. Mlodinow, ``Quantization of
electrodynamics in nonlinear dielectric media,'' Phys. Rev. A \textbf{30},
1860 (1984).

\bibitem{Carmichael}H. J. Carmichael, \emph{Statistical Methods in
Quantum Optics 1} (Springer, Berlin 2002).

\bibitem{Wigner}E. Wigner, ``On the Quantum Correction For Thermodynamic
Equilibrium,'' Phys. Rev. \textbf{40}, 749 (1932).

\bibitem{Higher-order}P. D. Drummond, ``Fundamentals of higher order
stochastic equations,'' J. Phys. A \textbf{47}, 335001 (2014).

\bibitem{Husimi}K. Husimi, ``Some Formal Properties of the Density
Matrix,'' Proc. Phys. Math. Soc. Japan \textbf{22}, 264 (1940).

\bibitem{ZambriniQ}R. Zambrini, S. M. Barnett, P. Colet and M. San
Miguel, ``Non-classical behavior in multimode and disordered transverse
structures in OPO. Use of the Q-representation,'' Eur. Phys. J. D
\textbf{22}, 461 (2003).

\bibitem{Glauber} R. J. Glauber, ``Coherent and Incoherent States
of the Radiation Field,'' Phys. Rev. \textbf{131}, 2766 (1963); E.
C. G. Sudarshan, ``Equivalence of Semiclassical and Quantum Mechanical
Descriptions of Statistical Light Beams,'' Phys. Rev. Lett. \textbf{10},
277 (1963).

\bibitem{Santagiustina}M. Santagiustina, E. Hernandez-Garcia, M.
San-Miguel, A. J. Scroggie and G.-L. Oppo, ``Polarization patterns
and vectorial defects in type-II optical parametric oscillators,''
Phys. Rev. E, \textbf{65}, 036610 (2002).

\bibitem{Zambrini}R. Zambrini, A. Gatti, L. Lugiato and M. San Miguel,
``Polarization quantum properties in a type-II optical parametric
oscillator below threshold,'' Phys. Rev A \textbf{68}, 063809 (2003).

\bibitem{Izus}G. Izús, M. San Miguel and D. Walgraef, ``Polarization
coupling and pattern selection in a type-II optical parametric oscillator,''
Phys. Rev. E \textbf{66}, 36228 (2002).

\bibitem{Ref3}S. Longhi, ``Alternating rolls in non-degenerate optical
parametric oscillators,'' J. Mod. Opt, \textbf{43}, 1569 (1996);
G. J. de Valcárcel and E. Roldán and K. Staliunas, ``Cavity solitons
in nondegenerate optical parametric oscillation,'' Opt. Commun. \textbf{181},
207 (2000); K. Staliunas, ``Transverse Pattern Formation in Optical
Parametric Oscillators,'' J. Mod. Opt. \textbf{42}, 1261 (1995);
G.-L. Oppo, M. Brambilla, D. Camesasca, A. Gatti and L. A. Lugiato,
``Spatiotemporal Dynamics of Optical Parametric Oscillators,'' J.
Mod. Opt. \textbf{41}, 1151 (1994); S. Longhi, ``Spatial solitary
waves in nondegenerate optical parametric oscillators near an inverted
bifurcation,'' Opt. Commun. \textbf{149}, 335 (1998); S. Longhi and
A. Geraci, ``Swift-Hohenberg equation for optical parametric oscillators,''
Phys. Rev. A \textbf{54}, 4581 (1996).

\bibitem{LPVol9}E. M. Lifshitz and L. P. Pitaevskii, \emph{Statistical
Physics}, \emph{Part 2}, in Volume 9 of the Course of Theoretical
Physics, (Pergamon Press, Oxford 1980).

\bibitem{Werner} P. D. Drummond and I. K. Mortimer, ``Computer simulations
of multiplicative stochastic differential equations,'' J. Comp. Phys.
\textbf{93}, 144 (1991); M. J. Werner and P. D. Drummond, ``Robust
Algorithms for Solving Stochastic Partial Differential Equations,''
J. Comp. Phys. \textbf{132}, 312 (1997).

\bibitem{Software} G. R. Collecutt and P. D. Drummond, ``Xmds: eXtensible
multi-dimensional simulator,'' Comput. Phys. Commun. \textbf{142},
219 (2001); S. Kiesewetter, R. Polkinghorne, B. Opanchuk and P. D.
Drummond,``xSPDE: extensible software for stochastic equations'',
SoftwareX in press.

\bibitem{RealSH}G. Kozyreff and M. Tlidi, ``Nonvariational real
Swift-Hohenberg equation for biological, chemical, and optical systems,''
Chaos \textbf{17}, 037103 (2007); M. Tlidi, Paul Mandel and R. Lefever
``Localized structures and localized patterns in optical bistability,''
Phys. Rev. Lett. \textbf{73}, 640 (1994).

\bibitem{Yin}J. Yin and D. P. Landau, ``Phase diagram and critical
behavior of the square-lattice Ising model with competing nearest-neighbor
and next-nearest-neighbor interactions,'' Phys. Rev. E \textbf{80},
051117 (2009).

\bibitem{GaussianApproxRef}Shang-Keng Ma, \emph{Modern theory of
critical phenomena}, (W. A. Benjamin, Reading, Massachusetts 1976);
P. Kopietz, L. Bartosch and F. Schtz, \emph{Introduction to the Functional
Renormalization Group} (Springer, Berlin, 2010).

\bibitem{StaliunasR2}K. Staliunas, ``Spatial and temporal noise
spectra of spatially extended systems with order disorder phase transitions,''
Int. Journal of Bifurcation and Chaos \textbf{11}, 2845 (2001); K.
Staliunas, ``Spatial and temporal spectra of noise driven stripe
patterns,'' Phys. Rev. E \textbf{64}, 066129 (2001).

\bibitem{StaliunasHypCSHEq}K. Staliunas and M. Tlidi, ``Hyperbolic
Transverse Patterns in Nonlinear Optical Resonators,'' Phys. Rev.
Lett. \textbf{94}, 133902 (2005).

\bibitem{StaliunasExpCSHE}K. Staliunas, G. Slekys and C. O. Weiss,
``Nonlinear Pattern Formation in Active Optical Systems: Shocks,
Domains of Tilted Waves, and Cross-Roll Patterns,'' Phys. Rev. Lett.
\textbf{79}, 2658 (1997).

\bibitem{StaliunasBook}K. Staliunas and V. J. Sanchez-Morcillo, \emph{Transverse
Patterns in Nonlinear Optical Resonators}, Springer Tracts Mod. Phys.
(Springer-Verlag, Berlin, 2003).

\bibitem{Gradshtein} I. S. Gradshteyn and I. M. Ryzhik. \emph{Table
of integrals, series, and products (7th ed)}. Academic Press, Amsterdam,
(2007).

\bibitem{PM_PRA37} P. D. Drummond and M. D. Reid, ``Laser bandwidth
effects on squeezing in intracavity parametric oscillation,'' Phys.
Rev. A \textbf{37}, 1806 (1988).\end{thebibliography}


\subsection*{Funding}

Australian Research Council (ARC); Conselho Nacional de Desenvolvimento
Científico e Tecnológico (CNPq).

\end{document}